\documentclass[sigconf]{acmart}
\AtBeginDocument{%
  \providecommand\BibTeX{{%
    \normalfont B\kern-0.5em{\scshape i\kern-0.25em b}\kern-0.8em\TeX}}}

\copyrightyear{2022}
\acmYear{2022}
\setcopyright{acmlicensed}\acmConference[WSDM '22]{Proceedings of the
Fifteenth ACM International Conference on Web Search and Data
Mining}{February 21--25, 2022}{Tempe, AZ, USA}
\acmBooktitle{Proceedings of the Fifteenth ACM International Conference on
Web Search and Data Mining (WSDM '22), February 21--25, 2022, Tempe, AZ,
USA}
\acmPrice{15.00}
\acmDOI{10.1145/3488560.3498410}
\acmISBN{978-1-4503-9132-0/22/02}

\usepackage{amsmath}
\usepackage{multirow}
\usepackage{multicol}
\usepackage{pifont}
\usepackage{enumitem}
\usepackage{subcaption} 
\usepackage{pgfplots}
\usepackage[noend]{algpseudocode}
\usetikzlibrary{pgfplots.groupplots}
\pgfplotsset{width=4cm,compat=1.9}

\newcommand{\xmark}{\ding{55}}

\newtheorem{theorem}{Theorem}

\usepackage[ruled, linesnumbered]{algorithm2e}
\begin{document}

% \title{Predicate Logic Graph Network}
% \title{Graph Logic Network for Relational Reasoning}
% \title{Graph Collaborative Reasoning for Recommendation and Link Prediction}
% \title{\mbox{Graph Logic Reasoning for Recommendation and Link Prediction}}
% \title{Graph Collaborative Reasoning based on Neural Logic}
% \title{Graph Convolutional Reasoning}
\title{Graph Collaborative Reasoning}

\author{Hanxiong Chen}
 \affiliation{
  \institution{Rutgers University}
  \institution{New Brunswick, NJ, US}
  \country{}}
 \email{hanxiong.chen@rutgers.edu}

\author{Yunqi Li}
 \affiliation{
  \institution{Rutgers University}
  \institution{New Brunswick, NJ, US}
  \country{}}
 \email{yunqi.li@rutgers.edu}
 
\author{Shaoyun Shi}
 \affiliation{
  \institution{Tsinghua University}
  \institution{Beijing, China}
  \country{}}
 \email{shisy17@mails.tsinghua.edu.cn}

\author{Shuchang Liu}
 \affiliation{
  \institution{Rutgers University}
  \institution{New Brunswick, NJ, US}
  \country{}}
 \email{shuchang.liu@rutgers.edu}
 
\author{He Zhu}
 \affiliation{
  \institution{Rutgers University}
  \institution{New Brunswick, NJ, US}
  \country{}}
 \email{hz375@cs.rutgers.edu}
 
\author{Yongfeng Zhang}
 \affiliation{
  \institution{Rutgers University}
  \institution{New Brunswick, NJ, US}
  \country{}}
 \email{yongfeng.zhang@rutgers.edu}


\begin{abstract}
Graphs can represent relational information among entities and graph structures are widely used in many intelligent tasks such as search, recommendation, and question answering. However, most of the graph-structured data in practice suffers from incompleteness, and thus link prediction becomes an important research problem. Though many models are proposed for link prediction, the following two problems are still less explored: (1) Most methods model each link independently without making use of the rich information from relevant links, and (2) existing models are mostly designed based on associative learning and do not take reasoning into consideration. With these concerns, in this paper, we propose Graph Collaborative Reasoning (GCR), which can use the neighbor link information for relational reasoning on graphs from logical reasoning perspectives. We provide a simple approach to translate a graph structure into logical expressions, so that the link prediction task can be converted into a neural logic reasoning problem. We apply logical constrained neural modules to build the network architecture according to the logical expression and use back propagation to efficiently learn the model parameters, which bridges differentiable learning and symbolic reasoning in a unified architecture. To show the effectiveness of our work, we conduct experiments on graph-related tasks such as link prediction and recommendation based on commonly used benchmark datasets, and our graph collaborative reasoning approach achieves state-of-the-art performance. 
\end{abstract}





\maketitle

\section{Introduction}
% Graph is a powerful data structure to hold information. GNN is a powerful tool to address graph related tasks. Many research work on GNN. 
Graph is able to describe the entities and their relations in many real-world systems and research problems, such as e-commerce user-item interactions, social networks, citation networks and knowledge graphs. Though graphs can encode rich relationships among plenty of entities, they still suffer from incompleteness \cite{wang2020h2kgat,rossi2020knowledge}. This issue gives rise to the link prediction task, which is to learn representations from the known data and then predict the potential valid connections. Link prediction is essential to many tasks such as knowledge graph reasoning, entity search, recommender systems and question answering.

Recent years have witness the success of knowledge graph embedding methods for link prediction \cite{bordes2013translating,wang2014knowledge,yang2015embedding,trouillon2016complex,dettmers2018convolutional}. The basic idea is to encode the entities and their relations into a low dimensional vector space while the inherent structure information of the graph is preserved. However, one drawback of these embedding-based models is that they usually process each \textit{(entity, relation, entity)} triplet independently without explicitly considering the information from neighborhood links, though information from neighbourhood nodes is considered. As a result, these methods are not able to capture the rich information from the neighbor connections and hence result in less informative embeddings \cite{arora2020survey, nathani2019learning}.

Another line of research is graph neural networks (GNNs), which have shown the power in many graph-related problems~\cite{hamilton2017inductive,Kipf:2016tc,veli2018graph}. These approaches are able to learn effective entity representations by aggregating its own representation and the representations of surrounding neighbors. The nodes in the graph can exchange information through message passing~\cite{gilmer2017neural}, which alleviates the problem of aforementioned embedding-based methods. Despite that GNNs could capture more information than those shallow embedding-based models, their key idea for handling link prediction tasks are actually similar---they aim to learn embeddings to capture the similarity patterns among entities, so that link prediction can be conducted by calculating the similarity for a pair of nodes over a specific relation. However, most GNN approaches are designed from a perceptual perspective and they seldom consider the logical relationship among entities and links for relational reasoning.

Logical reasoning is an essential and many times a natural way to conduct reasoning on graphs for two reasons. First, many triplets in the graph may be logically related and can be modeled together through logical connections. Take knowledge graph for example, the triplet (\textit{x, capitalOf, y}) logically implies the relation (\textit{x, locatedIn, y}). Thus, we can use implication operations in predicate logic to describe this connection between the two triplets as $(\textit{x, capitalOf, y})$ $\rightarrow$ $(\textit{x, locatedIn, y})$. The logical relationship among triplets, if accurately captured, would be helpful for predicting unknown links. Second, each triplet can be naturally represented as a predicate in logical reasoning, which makes it easy to model the link prediction task as a reasoning process. For example, we can treat the target triplet (\textit{x, locatedIn, y}) as a predicate expression \textit{locatedIn(x,y)}. Then, the link prediction task can be formulated as answering whether the logical expression $\textit{capitalOf(x,y)} \rightarrow \textit{locatedIn(x,y)}$ is true, given that the predicate \textit{capitalOf(x,y)} is true. If the logical expression is true, then we can infer that the target predicate should be true. In other words, the target triplet is a valid link.

In this paper, we explore an approach that transforms the link prediction task into a logical reasoning process on graphs. Our goal is to model the structure of a graph as simple Horn clauses so that link prediction can be conducted via logical reasoning. Inspired by \cite{shi2020neural,chen2020neural}, we apply modularized logical neural networks to learn the logical operations. Instead of using explicit hand-crafted logic rules as many previous approaches did, we introduce a method to convert graph structures into Horn clauses as potential rules to be learnt. The logical relations can be captured by the neural networks so that relational reasoning can be conducted on graphs. 
% The effectiveness of similar designs on real-world data have been shown in \cite{shi2020neural} and \cite{chen2020neural}. 

Technically, we propose a Graph Collaborative Reasoning (GCR) framework for relational reasoning over graphs. Specifically, we consider that links (or triplets) are potentially related to each other if they are connected by shared nodes. Based on this, we can infer a link through its neighbor links for relational reasoning. To compute the Horn clauses via deep neural networks, we encode each triplet as a predicate embedding, i.e., each entity in a given triplet is represented as a vector embedding and each relation is modeled as a neural module to encode the triplet. With the encoded predicate embeddings, we can construct the network structure using the neural modules in accordance with the modeled Horn clauses. The key benefits of our design compared to previous works are four aspects. First, we can take advantage of GNN strategies to aggregate rich information from neighbor links through message passing to make link predictions.
Second, we consider logical reasoning for link prediction, which can make use of the logical relationships between links.
Third, we incorporate logical reasoning without manually predefined rules, which makes our method easily adaptable to different scenarios. Finally, our model can handle uncertainty in logical reasoning. Our contributions can be summarized as follows:
\vspace{-1ex}
\begin{itemize}
    \item We introduce a new view of the link prediction task from logical reasoning perspectives. In this way, the link prediction task is translated into a true/false evaluation problem of predicate logical expressions.
    \item We propose the Graph Collaborative Reasoning (GCR) model, which conducts relational reasoning by taking advantage of the neighbor link information for message passing. 
    \item We show the effectiveness of our approach on various graph relational reasoning tasks on several real-world graph datasets.
\end{itemize}

In the following, we will present related works in Section \ref{sec:related work}. After that, in Section~\ref{sec:problem_formulation}, we formalize the link prediction task in logical language. Section~\ref{sec:Model} presents the details of our model and Section~\ref{sec:experiment} gives our experimental setup and results. We will conclude this work with outlooks for future work in Section~\ref{sec:conclusion}.
\vspace{-3ex}
\section{Related Works}
\label{sec:related work}
Existing techniques for link prediction can be roughly classified into three categories: translation-based, tensor factorization-based, and neural network-based. The translation-based models \cite{bordes2013translating,lin2015learning, wang2014knowledge, ji2015knowledge,yang2019transms} translate a head embedding into a tail embedding via a relation. The scoring function is defined as the distance between the translated head embedding and the tail embedding. Tensor factorization-based methods, such as RESCAL~\cite{nickel2011three}, ComplEx~\cite{trouillon2016complex}, RotatE~\cite{sun2019rotate}, DistMult~\cite{yang2015embedding} and HolE~\cite{nickel2016holographic}, consider the graph as a 3D adjacency matrix, which represents the head, tail and relation embeddings along each dimension. They apply operations such as linear mapping (RotatE), bilinear mapping (DistMult and ComplEx) or circular correlation operation (HolE) to obtain low-dimensional representations for each entity and relation. The deficiency of these methods lie in treating each triplet independently and thus the rich structural information in the graph cannot be adequately used.

Neural network-based methods, such as CNN-based~\cite{dettmers2018convolutional, nguyen2018novel} and GNN-based~\cite{schlichtkrull2018modeling, van2018graph} methods, use neural network structures to capture the rich information among the links. CNN-based methods, such as ConvE~\cite{dettmers2018convolutional}, use 2D convolution layers to extract the relationships between head entity embeddings and relation embeddings. The relations are represented as multiple feature maps, which are obtained through various filters. Then all these feature maps are concatenated and fed into a fully connected layer to get the projected embeddings for similarity calculation with the tail entity embeddings. These models still consider each triplet independently which also suffer from the aforementioned problem. 
GNN-based models, such as GCN~\cite{Kipf:2016tc}, GAT~\cite{veli2018graph} and GraphSAGE~\cite{hamilton2017inductive}, can help to resolve this issue by using message passing strategy to aggregate information from neighbor nodes so as to enrich the vector representation of each entity. Since the original design of these models are based on homogeneous graphs, they are unable to handle multi-relational link prediction tasks. Later, an extension of GCN named R-GCN~\cite{schlichtkrull2018modeling} is proposed to deal with multi-relational data. However, none of the above methods consider the logical relationships between nodes/links in the graph for relational reasoning. 

Recently, there have been some research works on integrating logic into link prediction. The related approaches can be broadly classified into hard-logic-based and soft-logic-based methods. The hard-logic-based methods focus on applying hard logic rules to the learning process \cite{demeester2016lifted,guo2016jointly,rocktaschel2015injecting,wang2018embedding,wang2015knowledge}. The problem of using hard logic rules is that the model does not tolerate to any violation. As a result, the logic rules need to be carefully designed and the application scenarios can be limited. For example, the hard-rule-based methods are able to handle rules like ``$x$ is the capital of $y$ implies $x$ is located in $y$,'' however, they can hardly deal with rules like ``user purchased a cellphone $x$ implies that user probably will purchase a phone case $y$,'' since the rule can be violated in some cases. 

To solve the problem, soft-logic-based methods try to handle this uncertainty by using soft logic constraints, which assign probabilities to the logic rules to make the model more tolerate to exceptions \cite{guo2018knowledge,zhang2019iteratively,qu2019probabilistic, ren2020beta, guo2020knowledge, harsha2020probabilistic, Zhang2020Efficient}. One powerful model is pLogicNet~\cite{qu2019probabilistic}, which is based on Markov Logic Network. It can learn the weight for each predefined logic rule to handle uncertainty and noise. 
% However, the high time complexity of inference over the rule structures limits its adaptability on large-scale graphs.
% Another recent work SLRE~\cite{guo2020knowledge} proposed a soft logic rule-based method, which applies logic rule constraints on relation embeddings without groundings and thus improves the scalability. 
However, these models usually need to ground the logic rules by traversing all potential valid links in a graph, which makes these methods difficult to scale to large graphs. Though recent works try to get rid of the grounding process by directly adding rule-based constraints on the relation vector representations \cite{ding2018improving,minervini2017regularizing,guo2020knowledge}, they can only deal with simple rules such as $(x, hypernym, y)\rightarrow (y, hyponym, x)$. 

% Other than these non-logical reasoning based methods, there are some recent works combine the link prediction tasks with hard logical rules~\cite{wang2015knowledge,guo2016jointly, wang2018embedding}. However, hard rules requires extensive human effort and expert knowledge to adapt to the task domain. This deficiency limits scalability of these methods since real world data is full of uncertainty which may even lead to the contradictory. Then soft rules based methods~\cite{qu2019probabilistic, ren2020beta, guo2020knowledge} are proposed to handle such uncertainties.
All of the aforementioned logical rule-based methods need explicitly predefined logic rules either as part of a pipelined framework or as a constraint of the learning process. This makes the model highly dependent on the effectiveness of the predefined logic rules. An open challenge, as mentioned in \cite{guo2020knowledge}, is to design models that can handle not only simple (manually) designed rules but also complex learned rules while considering the scalability and uncertainty. Although soft-logic-based methods can be more flexible than hard-logic-based approaches, these works all need the background knowledge of the data so that logical rules can be created reasonably, which needs considerable manual efforts.

% These simple rules are actually special cases of Horn clause, which is a flexible subset of predicate logic that can express higher-order connectivity patterns among relations. 

\section{Problem Formulation}
\label{sec:problem_formulation}
The link prediction task predicts the potential connections among nodes/entities from the known information in a graph. Different from previous works which treat each triplet independently, we consider that triplets may have potential relations to each other if they have shared nodes. This information is helpful in many cases. For example, in a social network, the reason that Alice and Bob follow each other is probably because of their common habits. That means the triplet $(\textit{Alice, follows, Bob})$ is valid due to $(\textit{Alice, likes, Pop})$ and $(\textit{Bob, likes, Pop})$, which can be represented as the logical expression $\textit{likes}(\textit{Alice, Pop})\wedge\textit{likes}(\textit{Bob, Pop})\rightarrow\textit{follows}(\textit{Alice, Bob})$. Based on this, we can take advantage of the neighbor information to help link prediction. To realize this idea, we model the link prediction task in three steps: 1) convert the graph structure into a logic expression; 2) use neural modules to encode triplets as predicate embeddings; 3) apply logical constrained modules to generate ranking scores. The details for step 2) and 3) will be given in Section~\ref{sec:Model}. In this section, we focus on how to convert an graph structure into a logic expression and how to formulate the link prediction task as a true/false evaluation problem of logical expressions.

Suppose we have a graph $\mathcal{G}=(\mathcal{V}, \mathcal{R}, \mathcal{T})$, where $\mathcal{V}$ is the vertex set, $\mathcal{R}$ is the relation set, and the known triplets (edges) in the graph are represented as $\mathcal{T}$. For any $v_i, v_j\in \mathcal{V}$ and a relation $r_k\in \mathcal{R}$, we need to predict if the target triplet $T_x = (v_i, r_k, v_j)$ is valid, where $T_{x}\notin \mathcal{T}$. To solve this problem, we first get the neighbors of both $v_i$ and $v_j$ and get all the triplets $\mathcal{T}_{ij}$ that contain either $v_i$ or $v_j$. 
\begin{equation}
\begin{aligned}
\mathcal{T}_{ij}&=\{(v_i, r_{in}, v_n)|v_n\in \mathcal{N}_i\} \cup \{(v_j, r_{jm}, v_m)|v_m\in \mathcal{N}_j\}\\
&=\{r_{in}(v_i, v_n)|v_n\in \mathcal{N}_i\} \cup \{r_{jm}(v_j, v_m)|v_m\in \mathcal{N}_j\}
\end{aligned}
\end{equation}
where $\mathcal{N}_i$ and $\mathcal{N}_j$ are the neighbor vertex sets of node $v_i$ and $v_j$, respectively, and the link is considered as a predicate. 
% Since not all the triplets $\mathcal{T}_{ij}$ may contribute to the prediction of the target triplet $T_{x}$, 
Since it is possible that not all the triplets in $\mathcal{T}_{ij}$ are the reasons of the target triplet $T_{x}$, we apply the OR operator to model the prediction task. 
% It can be interpreted as that: as long as any neighbor or combination of neighbours can imply the target relation, then this target relation can be valid. 
The intuition here is that: the reason that $T_{x}$ holds could be any of its neighbour links or any combination of its neighbour links.
We translate this idea into the following expression:
\begin{equation}
% \hspace{-20pt}
\label{eq:1}
\begin{split}
    &(T_1\rightarrow T_{x})\vee (T_2\rightarrow T_{x}) \vee \cdots \vee (T_n\rightarrow T_{x}) \\ 
    \vee~ &(T_1 \wedge T_2 \rightarrow T_{x})\vee (T_1 \wedge T_3 \rightarrow T_{x})\vee \cdots \vee(T_{n-1}\wedge T_n\rightarrow T_{x}) \\
    \vee~ &(T_1 \wedge T_2 \wedge T_3 \rightarrow T_{x})\vee \cdots \vee (T_{n-2} \wedge T_{n-1}\wedge T_n\rightarrow T_{x}) \\
    &\cdots\\
    \vee~ & (T_1\wedge T_2 \wedge\cdots \wedge T_n \rightarrow T_{x})
\end{split}
\end{equation}
where $T_1,T_2\cdots T_n$ are triplets in $\mathcal{T}_{ij}$, and ``$\rightarrow$'' is called the implication operation\footnote{In classical logic, $p\rightarrow q$ is equivalent to $\neg p \vee q$}. 
This expression contains not only simple Horn clauses, such as $(T_1\rightarrow T_{x})$, but also higher-order Horn clauses, such as $(T_1 \wedge T_2 \rightarrow T_{x})$ and $(T_1\wedge T_2 \wedge\cdots \wedge T_n \rightarrow T_{x})$. 
Based on this definition, we have the following theorem:
\begin{theorem}
Equation \eqref{eq:1} is true if and only if $T_x$ is true.
\end{theorem}

To show why, we first have the following lemma:
\begin{lemma}
Let the premise $p$ be true, then the clause $p\rightarrow q$ is true if and only if the conclusion $q$ is true.
\end{lemma}

The lemma naturally follows from the definition of the implication operation: $p\rightarrow q \Leftrightarrow \neg p \vee q$. Now back to Theorem 1, since all of the known triplets in the training data are valid, we know that each $T_\ast\in \mathcal{T}_{ij}$ is true, and thus any conjunction among $T_\ast$ is also true. As a result, if $T_x$ is true, then Eq.\eqref{eq:1} must be true, and if Eq.\eqref{eq:1} is true, we know that at least one of the Horn clauses in Eq.\eqref{eq:1} must be true, and thus $T_x$ must be true,
% Then we know that if any of these Horn clauses is true, then the Expression in Eq.\eqref{eq:1} will be true, 
meaning that $T_{x}$ is a valid triplet. 
% This conclusion can be derived from the following theorem:
Now the problem of judging if a target triplet $T_{x}$ is valid or not becomes answering the question that whether the logic expression in Eq.\eqref{eq:1} is true given the known triplets. The intuition here is that $T_x$ is true as long as at least one of its known neighbour connections or their conjunctions can imply $T_x$.

However, one problem is that the size of the expression is huge, which is equal to $O(2^n)$---the size of the power set of $\mathcal{T}_{ij}$, making it impractical to implement Eq.\eqref{eq:1}. Fortunately, we can simplify the expression in Eq.\eqref{eq:1} through implication rule and De Morgan's Law\footnote{De Morgan's Law, in formal language, is written as $\neg(p\vee q) \Leftrightarrow \neg p \wedge \neg q$ and $\neg(p\wedge q) \Leftrightarrow \neg p \vee \neg q$}, which translates Eq.\eqref{eq:1} into following simplified form:
\begin{equation}
\label{eq:2}
    \neg T_1 \vee \neg T_2 \vee \cdots \vee \neg T_n \vee T_{x}
\end{equation}

Compare to the $O(2^n)$ complexity of Expression \eqref{eq:1}, the complexity of Expression~\eqref{eq:2} is only $O(n)$. We will use Expression~\eqref{eq:2} for our model implementation. 
%Figure~\ref{fig:overview} gives an example of the conversion process. 
In the next section, we will introduce how to encode triplets into embeddings and then build logic neural networks to generate ranking scores for relational reasoning.

\begin{figure}[t!]
\centering
\includegraphics[scale=0.5]{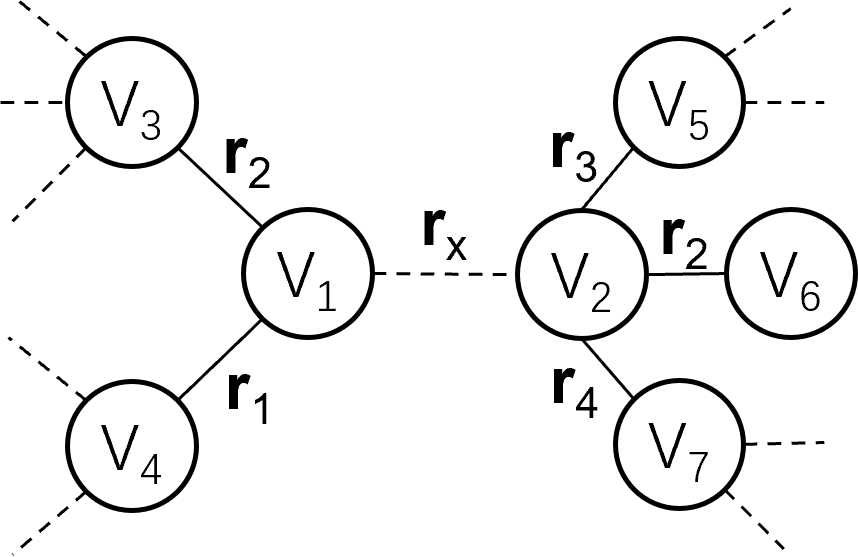}
\vspace{-10pt}
\caption{An example of link prediction on a heterogeneous graph. From a logical view, $r_x(v_1, v_2)$ to be true could result from any order of combinations of the neighbor links, e.g. first-order $r_1(v_1, v_4)$, second-order $r_1(v_1, v_4)\wedge r_2(v_2, v_6)$ or even higher-order $r_1(v_1,v_4)\wedge r_2(v_1,v_3)\wedge \ldots \wedge r_3(v_2,v_5)$.}
\vspace{-15pt}
\label{fig:heto_graphs}
\end{figure}

\section{Graph Collaborative Reasoning}
\label{sec:Model}
% \begin{figure*}[t]
%     \centering
%     \includegraphics[scale=0.6]{Figure/JPEG/overview.png}
%     \caption{The transformation from (a) a regular heterogeneous graph to (b) the tree spanned from vertex v1, then to (c) the predicate logic graph spanned from the predicate $r_2(v_1, v_2)$.}
%     \label{fig:overview}
% \end{figure*}
% A graph includes vertices and edges. In many application scenarios, vertices are used to represent the entities, such as users in social networks or recommender systems, while the edges are the relations between entities. From the logic perspective, the connection between two vertices in a graph can be expressed as a predicate, and then the link prediction task can be converted to a reasoning problem, that is to decide if the predicate is true or false. This has been explained in Section~\ref{sec:problem_formulation}. 

% Since the links between vertices can be converted to predicates, we can easily transform a given graph 
% % (vertices as variables and edges as relations) 
% into a logical graph, where vertices are the logical variables while the edges represent the predicate operations over the variables. In other words, 
Our GCR framework views a graph from the edge perspective and aims to learn the relationship between adjacent edges that are connected by common nodes. 
% reflect the logical relations among triplets. 
Instead, traditional GNN views a graph from the node perspective and aims to learn the relationship between nodes that are connected by common edges. 
% focus on exploring the relations among vertices. 
In Figure~\ref{fig:heto_graphs}, we use an example to show how a link prediction task on a heterogeneous graph can be viewed from logical perspective. In this example, we hope to predict if node $v_1$ and $v_2$ could be connected by relation $r_x$. Intuitively, $r_x(v_1,v_2)$ could be true due to: 1) any first-order implication, e.g. $r_1(v_1,v_4)\rightarrow r_x(v_1, v_2)$ or $r_3(v_2,v_5)\rightarrow r_x(v_1, v_2)$ is true, or 2) any second-order implication, e.g. $r_1(v_1,v_4)\wedge r_2(v_2,v_6)\rightarrow r_x(v_1, v_2)$ is true, or even higher-order implication, e.g. $r_1(v_1,v_4)\wedge r_2(v_1,v_3)\wedge \ldots \wedge r_3(v_2,v_5)\rightarrow r_x(v_1, v_2)$ is true. With Eq.\eqref{eq:2}, this problem can be simplified as predicting if the following expression consisting of all neighbour links is true:
\begin{equation}
    \neg r_1(v_1,v_4) \vee \neg r_2(v_1,v_3) \vee \neg r_2(v_2,v_6) \vee \neg r_3(v_2,v_5) \vee \neg r_4(v_2,v_7) \vee r_x(v_1,v_2)
\end{equation}

In the following subsections, we will show the details of our graph collaborative reasoning framework.

\subsection{Node and Link Encoding}
We treat each type of relation in the graph as a predicate, e.g., each of the previously mentioned relations such as $capitalOf$, $locatedIn$, $follows$, $likes$ is a predicate. We learn each node as a vector embedding, same as traditional graph neural networks. Meanwhile, we learn each predicate (relation type) as a small neural module. The predicate serves as a function that converts the two connected nodes into a latent vector in the reasoning space, e.g., to process the link $(Alice,likes,Pop)$, we write it as the predicate form $likes(Alice,Pop)$, then the node embeddings of $Alice$ and $Pop$ are fed into the neural module of $likes$ to get the output representation for this link.
% Different predicates, in our framework, are represented as different neural modules. For example, 
% We apply a multi-layer perceptron (MLP) with non-linear activation function to map the concatenation of the latent vectors of the head and tail entities into a logic space using the predicate neural module.
More specifically, the encoding process is given as:
\begin{equation}\label{eq:encoding}
    \mathbf{e}_{h,t}^r = P_r(\textbf{e}_h, \textbf{e}_t) = \mathbf{W}_2^r\phi(\mathbf{W}_1^{r}(\mathbf{e}_h ; \mathbf{e}_t) + \mathbf{b}_1^r) + \mathbf{b}_2^r
\end{equation}
where $P_r(\cdot, \cdot)$ is the predicate function for relation $r\in \mathcal{R}$; $\textbf{e}_h, \textbf{e}_t \in \mathbb{R}^d$ are embeddings for head and tail entities; $(\cdot ; \cdot)$ is concatenation operation; $\phi(\cdot)$ is ReLU activation function; $\textbf{W}_1^r, \textbf{W}_2^r \in \mathbb{R}^{n\times 2d}$ and $\textbf{b}_1^r, \textbf{b}_2^r \in \mathbb{R}^n$ are network parameters and bias terms. Here $\mathbf{e}_{h,t}^r$ is the predicate embedding of the triplet $(v_h, r, v_t)$. One thing we need to clarify here is that the order of the head and tail entity embeddings must be correctly sorted during the implementation, because we use concatenation operation to combine the head and tail embeddings, different ordering of head and tail concatenation will result in different outputs. However, this can be a problem for undirected graphs where the triplet $(h, r, t)$ should have the same vector representation as $(t, r, h)$. In our implementation, we solve this problem by assigning a unique ID to each vertex in the graph and sort their ID in ascending order. This will make sure that the triplet always comes with the smaller ID entity as the head entity while the bigger ID entity as the tail entity. For directed graphs, we will not conduct the sorting operation since the ordering is part of the graph information. 

\subsection{Logical Reasoning Modules}

After obtaining all the encoded triplet vectors, we can rewrite the Expression~\eqref{eq:2} in the predicate embedding form:
\begin{equation}
\label{eq:logic_exp}
    (\neg \mathbf{e}_{i, n_1}^{r_{i n_1}} \vee \neg \mathbf{e}_{i, n_2}^{r_{i n_2}}\vee \ldots \vee \neg \mathbf{e}_{j, m_1}^{r_{j m_1}} \vee \neg \mathbf{e}_{j, m_2}^{r_{j m_2}}) \vee \mathbf{e}_{i, j}^{r_{x}}
\end{equation}
Here $\mathbf{e}_{i, j}^{r_{x}}$ represents the predicate embedding for the target triplet $T_{x}=(v_i, r_x, v_j)$. Since the target triplet is unknown and need to be predicted, we use $r_x$ instead of $r_{i,j}$ to make the notation concise. $\mathbf{e}_{i,n_k}^{r_{i n_k}}$ and $\mathbf{e}_{j,m_k}^{r_{j m_k}}$ are the encoded predicate embeddings for the known neighbour triplets in the graph that contain either $v_i$ or $v_j$. Our goal is to predict if the above logical expression is true in a continuous reasoning space. We define a constant vector $\textbf{T}$, which is an anchor vector in the reasoning space that represents true. It is randomly initialized and kept unchanged during model training. We expect that the final vector representation of the entire expression is close to this true vector $\mathbf{T}$ if the target triplet $T_{x}$ is valid. Otherwise, the vector representation of the logical expression should be far from $\mathbf{T}$.

To achieve this goal, we create neural modules $\text{OR}(\cdot,\cdot)$ and $\text{NOT}(\cdot)$ to represent the logical operations $\vee$ and $\neg$, where each module is an MLP with ReLU as activation function. To allow the neural logical modules to perform logical operations as expected, we add logical regularizers to the neural modules to constrain their behavior as defined in \cite{shi2020neural,chen2020neural}. The regularizers are not only added to the input predicate embeddings but also to the intermediate hidden vectors as well as the output vector to guarantee that all the embeddings are in the same representation and reasoning space. The logic constraint is represented as $\mathcal{L}_{logic}$. 

% \begin{figure}[t!]
%     \centering
%     \includegraphics[scale=0.55]{Figure/JPEG/network.png}
%     \caption{The network structure of the logic expression $\big(r_1(v_1, v_2) \rightarrow r_3(v_2, v_3)\big) \vee \big(r_2(v_3, v_4) \rightarrow r_3(v_2, v_3)\big) \vee \big(r_1(v_1, v_2) \wedge r_2(v_3, v_4) \rightarrow r_3(v_2, v_3)\big)$, which can be equivalently converted to $\neg r_1(v_1, v_2) \vee \neg r_2(v_3, v_4) \vee r_3(v_2, v_3)$ by De Morgan Law. The network is assembled using the converted logic expression.}
%     \label{fig:network}
% \vspace{-10pt}
% \end{figure}
\begin{figure}[t!]
    \centering
    \includegraphics[scale=0.5]{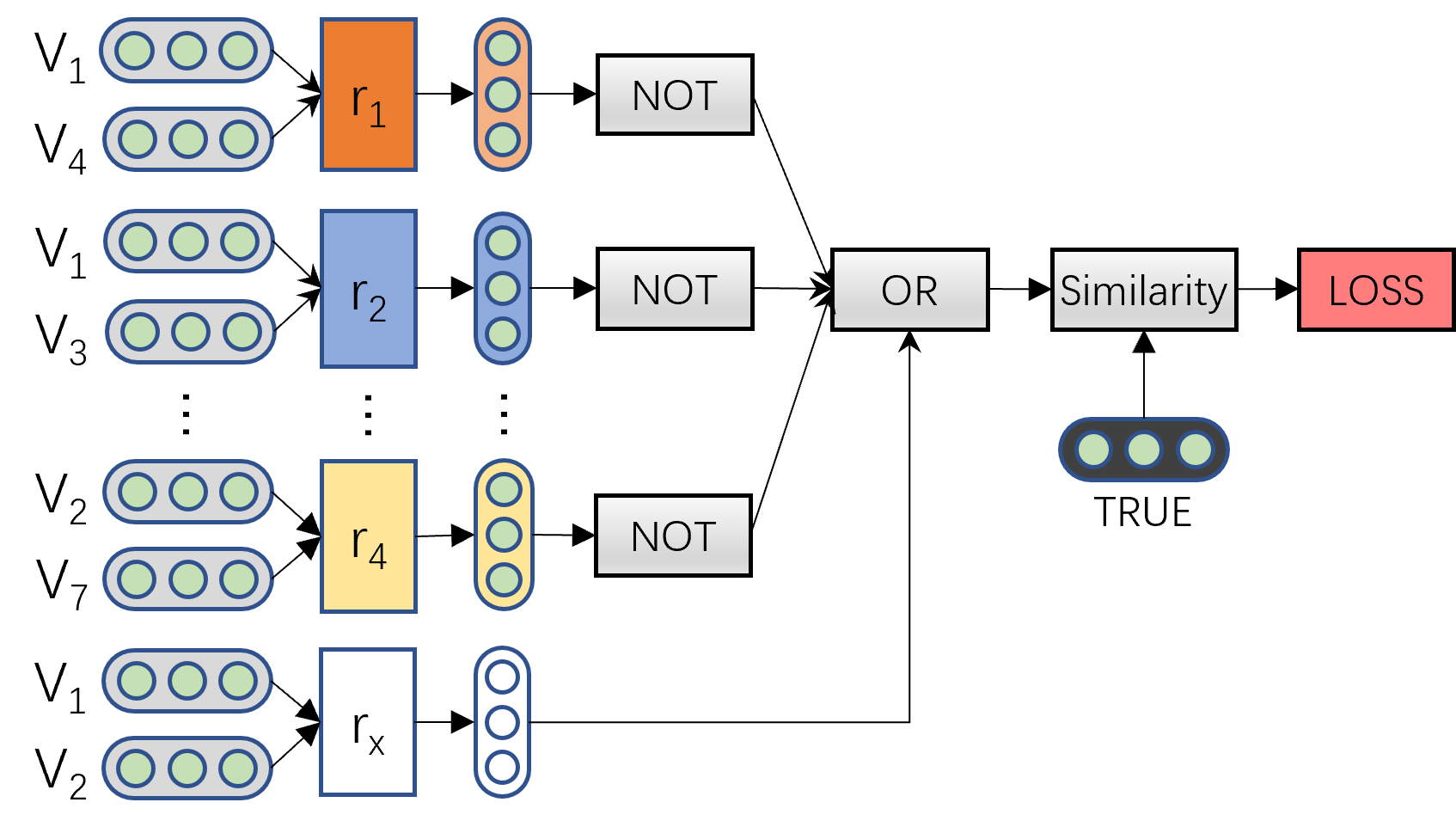}
    \caption{The logical network structure of the link prediction task given in Figure~\ref{fig:heto_graphs}. The network is assembled using the logical equivalent expression which is converted via De Morgan's Law.}
    \label{fig:network}
\vspace{-10pt}
\end{figure}

With these logical modules, we can then assemble a neural network for Expression~\eqref{eq:logic_exp}. To make the explanation easy to follow, we use a specific example as shown in Figure~\ref{fig:network} to explain the network construction process. This reasoning network structure is corresponding to the heterogeneous graph given in the Figure~\ref{fig:heto_graphs}. Suppose we are given two vertices $v_1$ and $v_2$, our goal is to predict if they could have a valid connection through relation $r_x$. According to the steps mentioned before, we need to first find the neighbors of both $v_1$ and $v_2$, in this example are $\{v_3,v_4,v_5,v_6,v_7\}$. Then we feed these vertex pairs into the corresponding predicate encoders to get the predicate embeddings based on Eq.\eqref{eq:encoding}. By sending these predicate embeddings into the $\text{NOT}(\cdot)$ module, we can calculate the negated embeddings, e.g. $\neg \textbf{e}_{1,4}^{r_1}$. After that, we follow the structure of Eq.\eqref{eq:logic_exp} to send the target predicate embedding $\textbf{e}_{1,2}^{r_x}$ together with the negated embeddings into the $\text{OR}(\cdot, \cdot)$ module to get the final vector representation of the entire expression in the reasoning space. Since $\text{OR}(\cdot, \cdot)$ only takes two inputs at one time, we calculate the joint embedding for more than two predicate embeddings in a recurrent manner. That is, we first send two predicates, e.g. $\neg \textbf{e}_{1,4}^{r_1}$ and $\neg \textbf{e}_{1,3}^{r_2}$ in Figure~\ref{fig:network}, into the $\text{OR}$ module and get the hidden vector $\textbf{e}^{r_1,r_2}$, which represents the result of $\neg \textbf{e}_{1,4}^{r_1} \vee \neg \textbf{e}_{1,3}^{r_2}$. The next predicate embedding in the expression and the previous hidden vector $\textbf{e}^{r_1, r_2}$ will be sent into the same OR neural module. This process is recurrently conducted until we get the final vector representation of the entire logical expression. However, we need to guarantee that the order information will not affect the final output since the logical OR operation need to satisfy the associativity and commutativity laws. This is done by randomly shuffling the order of the expression terms in each iteration. The following equations describe the process shown in Figure~\ref{fig:network}:
\begin{equation}
\label{eq:gln}
\begin{split}
    % \neg \mathbf{e}_{i,j}^{r_k} &= \text{NOT}(\textbf{e}_{i,j}^{r_k}), \forall i,j\in \mathcal{N}_i \cup \mathcal{N}_j\\
    % \neg \mathbf{e}_{i,n_x}^{r_{in_x}} &= \text{NOT}(\textbf{e}_{i,n_x}^{r_{in_x}}), \forall n_x\in \mathcal{N}_i\\
    \neg \mathbf{e}_{i,j}^{r_k} &= \text{NOT}(\textbf{e}_{i,j}^{r_k}), \forall i,j\\
    \textbf{E} &= \text{OR}\left(\neg\textbf{e}_{1,4}^{r_1},\neg\textbf{e}_{1,3}^{r_2},\cdots,\neg\textbf{e}_{2,7}^{r_4},\textbf{e}_{1,2}^{r_x}\right)
\end{split}
\end{equation}

For expressions that have more predicate embeddings in the expression, we can simply add more recurrent steps and do the same operation as mentioned above. The final output $\textbf{E}$ is the vector representation of the whole expression in the form of Eq.\eqref{eq:logic_exp}. The next step is to evaluate the distance between $\textbf{E}$ and the constant true vector $\textbf{T}$. As stated before, this true vector is randomly initialized and will not be updated during the learning process, as a result, it can be treated as an anchor vector in the reasoning space. Here we apply cosine similarity as the measure:
\begin{equation}
\label{eq:consine_sim}
    CosineSim(\textbf{E}, \textbf{T}) = \frac{\textbf{E}\cdot \textbf{T}}{\|\textbf{E}\|\|\textbf{T}\|}
\end{equation}
This cosine similarity measure is the score function and the output is treated as the ranking score to generate the entity ranking list.

\subsection{Learning Algorithm}

We use pair-wise learning algorithm~\cite{bpr} to train our model. Specifically, during the training process, for each known triplet in the training set, we fix the head entity and their corresponding relation and sample another entity as the tail. We treat expression created by this fake triplet $T_{x}^\prime$ as the negative sample. The same operation can be done one more time by holding the tail entity unchanged and replace the head entity. One thing need to mention here is that the neighbors to be sampled for creating the logic expression are never changed even when the head or tail entity is replaced, i.e., the only change in Eq.\eqref{eq:2} is to replace $T_{x}$ with $T_{x}^\prime$. The expression for the valid triplet, known as the positive sample, is evaluated based on Eq.\eqref{eq:consine_sim} and we have the score $s^+_T$, while the score for negative sample is $s^-_{T'}$. The loss function is written as:
\begin{equation}
\label{eq:gln_loss}
    \mathcal{L}_{gcr} = -\sum_{T\in \mathcal{T}, T^\prime\notin \mathcal{T}}\ln \sigma(\alpha(s^+_T - s^-_{T'}))
\end{equation}
where $\sigma(\cdot)$ is the logistic sigmoid function $\sigma(x)=\frac{1}{1+e^{-x}}$; $\alpha$ is an amplification coefficient, which is set to 10 in our implementation. We can apply an optimization algorithm to minimize $\mathcal{L}_{gcr}$ so as to maximize the distance between positive and negative samples. By integrating the logical regularizers into the graph collaborative reasoning network loss, we get the final loss function:
\begin{equation}
    \mathcal{L}=\mathcal{L}_{gcr} + \lambda_l\mathcal{L}_{logic} +\lambda_{\Theta}||\Theta||_2^2
\end{equation}
where $\lambda_l$ is the coefficient of the logical regularizers; $\Theta$ represents all the trainable parameters of the model, including entity embeddings, predicate encoder parameters and the parameters of the neural logical modules; $\lambda_{\Theta}$ is the $\ell_2$-norm regularization weight; We use back propagation~\cite{rumelhart1986learning} to optimize the model parameters. The pseudo-code for the entire training algorithm, including neighbor sampling, is given in Appendix~\ref{A:alg}.

\section{Experiments}
\label{sec:experiment}
In this section, we evaluate our proposed model on two types of link prediction tasks---graph link prediction and recommendation. The reason why we choose these two tasks for evaluation are based on two considerations: the uncertainty of the target links and the type of the graph structure. 

Knowledge graph is a type of heterogeneous graph that contains multi-type relations among entities, which makes the link prediction task challenging. It requires the model to predict not only if two entities will be connected but also determine which type of relation connects them. 
% For some well-known knowledge graphs such as Freebase~\cite{bollacker2008freebase}, 
The information in knowledge graphs is usually based on objective facts. That means each link can only be grounded as either true or false---not anything in between---since the links represent facts.
% neither will the link be true or false with probabilities.
Recommendation task usually considers a bipartite graph, which takes user and item as two types of nodes. The model needs to predict if a user and an item can be potentially connected so that we can recommend an item to a target user. The challenge is that the data is human generated which contains uncertainty and noise, so that it is usually not suitable to assign a deterministic truth value for a specific pair of nodes. 

As we mentioned before, our model can handle the uncertainty for relational reasoning over multi-relational graphs, we choose these two tasks to verify the effectiveness of our graph collaborative reasoning model by answering the following research questions:
% in the experiments:
\begin{itemize}
    \item \textbf{RQ1}: What is the performance of GCR in terms of graph link prediction and recommendation tasks? Does it outperform state-of-the-art models? (Section \ref{sec:result_1})
    \item \textbf{RQ2}: If and how does the logic regularizer help to improve the performance? (Section \ref{sec:result_2})
    \item \textbf{RQ3}: What is the impact of logical reasoning on few-shot data? (Section \ref{sec:result_3})
\end{itemize}

\begin{table}[t]
\caption{Statistics of the recommendation datasets.}\label{tb:dataset}
\vspace{-5pt}
\begin{tabular}{ccccc}
\toprule
Dataset& \#Users &\#Items &\#Interaction & Density\\
\midrule
Beauty &22,363 &12,101 &198,502 &0.073\%\\
Clothing &39,387 &23,033 &278,677 & 0.031\%\\
\bottomrule
\end{tabular}
\vspace{-10pt}
\end{table}

\begin{table*}[t]
    \caption{Baseline models used for either graph link prediction task or recommendation task.}
    \vspace{-5pt}
    \centering
    \begin{tabular}{cccccccccc}
    \toprule
        Baseline & TransE & DistMult & ConvE & R-GCN & pLogicNet &pGAT& BPR-MF & NCR & NGCF \\
        \midrule
        KG Completion & \checkmark & \checkmark & \checkmark & \checkmark & \checkmark & \checkmark & \xmark & \xmark & \xmark \\
        Recommendation & \checkmark & \checkmark & \checkmark & \xmark & \xmark & \xmark & \checkmark & \checkmark & \checkmark \\
    \bottomrule
    \end{tabular}
    \label{tb:baselines}
    % \vspace{-5pt}
\end{table*}

\subsection{Datasets}
For graph link prediction task, we use a well-known dataset \textbf{FB15k-237}~\cite{toutanova2015representing}, which is a subset of FB15k by removing the inverse relations in the training set to avoid data leakage. It contains 14,541 entities and 237 relations. The training dataset contains 272,115 edges while the validation and testing sets contain 17,535 and 20,466 edges, respectively. In the experiment, we use the same training, validation and testing data splits as described in~\cite{toutanova2015representing}.

For recommendation task, we use a publicly available Amazon e-commerce dataset \cite{mcauley2015image}, which includes the user, item and rating information. The user-item interaction matrix can be viewed as a bipartite graph with two types of nodes, i.e. user and item, and a single relation, which is the purchase relation in e-commerce scenario. This is a sparse dataset which makes personalized recommendation challenging.
% It covers 24 different categories, and 
We take \textbf{Beauty} and \textbf{Clothing} sub-categories for our experiments to explore both the link prediction performance and how our model performs in few-shot scenarios. Statistics of the datasets are shown in Table~\ref{tb:dataset}.

\subsection{Baselines}
We select several representative models for graph link prediction and recommendation to evaluate the performance of our proposed method. For graph link prediction, we use translation-based, tensor factorization-based, neural network-based as well as logic-based baselines for performance comparison.

\begin{itemize}
    \item \textbf{TransE}~\cite{bordes2013translating}: A classical translation-based knowledge graph embedding algorithm. The scoring function for each triplet is given as $\|\textbf{h}+\textbf{r}-\textbf{t}\|_p$, where $\textbf{h}, \textbf{r},\textbf{t}$ are entity and relation embeddings and $\|\cdot\|_p$ is the $p$-norm of the output vector.
    \item \textbf{DistMult}~\cite{yang2015embedding}: This is a tensor factorization-based knowledge graph embedding algorithm, which is a bilinear diagonal model.
    \item \textbf{ConvE}~\cite{dettmers2018convolutional}: This approach uses 2D-convolutional operation over embeddings to capture the information from the triplets, which is one of the state-of-the-art models on graph link prediction.
    \item \textbf{R-GCN}~\cite{schlichtkrull2018modeling}: This is a graph neural network based method, which extends Graph Convolutional Network (GCN) \cite{Kipf:2016tc}
    % , which originally designed for homogeneous graph, 
    to handle multi-relational link prediction tasks. 
    % \item \textbf{MLN}~\cite{richardson2006markov}: \textbf{M}arkov \textbf{L}ogic \textbf{N}etwork. This is a rule-based method which can handle uncertainty by learning weights to manually predefined logical rules.
    \item \textbf{pLogicNet}~\cite{qu2019probabilistic}: The Probabilistic Logic Network, which is a logic-based relational reasoning model. It defines the joint distribution of all possible triplets trough Markov Logic Network (MLN) with logic rules, so that the optimization process can be efficient.
    \item \textbf{pGAT}~\cite{harsha2020probabilistic}: This is a state-of-the-art MLN-based relational reasoning model, which combines MLN with graph attention network for link prediction.
    
\end{itemize}

For recommendation task, we also use the \textbf{TransE}, \textbf{DistMult} and \textbf{ConvE} knowledge graph embedding models as baselines since these models can also handle recommendation tasks. Other than that, we also use three recommendation models to explore if the GCR relational reasoning model can outperform those models that are specifically designed for recommendation, including:
\begin{itemize}
    \item \textbf{BPR-MF}~\cite{bpr}: This is a pair-wise ranking model for recommendation. We implement the prediction function under the BPR framework by following \cite{koren2009matrix}, which considers user, item and global bias terms for matrix factorization.
    % \item \textbf{NeuMF}~\cite{he2017neural}: This is a neural network based collaborative filtering algorithm by learning non-linear matching functions for recommendation.
    \item \textbf{NCR}~\cite{chen2020neural}: This is a state-of-the-art reasoning-based recommendation framework. It utilizes neural logic reasoning to model recommendation tasks.
    \item \textbf{NGCF}~\cite{wang2019neural}: This is an extension of GCN for recommendation task. It allows for multi-hop user-item information aggregation via message passing to enhance the user and item embeddings for recommendation.
\end{itemize}

We use Table~\ref{tb:baselines} to show which baseline model can be used for which link prediction task. For reproducibility, we present the details of the experimental setup for training and evaluating our model and baselines in Appendix~\ref{A:exp_setting}.

\subsection{Evaluation Protocol}
\subsubsection{\textbf{Link Prediction}} In the evaluation step, for each triplet, we first hold the head entity and replace the tail entity with ones that the head entity is not connected to. Then we do the same operation to hold the tail entity and replace the head entity. We call these generated non-existent triplets as negative samples. For each triplet and its corresponding negative samples, we calculate their evaluation metrics. The final results are averaged over all the triplets. We follow existing works~\cite{bordes2013translating,yang2015embedding} and use the filtered setting for evaluation. We report Mean Reciprocal Rank (MRR) and top-$K$ Hit rate (Hit@$K$) evaluation metrics in our results.

\subsubsection{\textbf{Recommendation}} In recommendation task, for each user-item interaction, we only sample items for each user that the user has never interacted with. Then these negative samples together with the target triplets constitute a user ranking list. Then we calculate the corresponding ranking score for each user and report the final scores by averaging over all the users. Here we use Normalized Discounted Cumulative Gain (NDCG@$K$) and Hit rate (Hit@$K$) metrics in our recommendation evaluation.

\begin{table*}[t]
\setlength{\tabcolsep}{3.5pt}
\caption{Link prediction performance on three datasets with metrics NDCG (N) and Hit Ratio (HR). We use underline (\underline{number}) to show the best result among the baselines, and use bold font to mark the best result of the whole column. We use star (*) to indicate that the performance is significantly better than all baselines. The significance is at 0.05 level based on paired $t$-test. The last row shows the relative improvement of our model against the best baseline performance.}\label{tb:results}
\vspace{-10pt}
\small
\centering\begin{tabular}{lccccccccccc}
\toprule
\multirow{2}{*}{} &\multicolumn{3}{c}{\textbf{FB15k-237}} & \multicolumn{4}{c}{\textbf{Beauty}} & \multicolumn{4}{c}{\textbf{Clothing}}\\ 
\cmidrule(lr){2-4}
\cmidrule(lr){5-8}
\cmidrule(lr){9-12}
&MRR &Hit@1 &Hit@3 &NDCG@5 &NDCG@10 &Hit@5 & Hit@10 &NDCG@5 &NDCG@10 &Hit@5 & Hit@10\\
\midrule
 TransE &0.326 &0.229 &0.363 &0.0063 &0.0086 &0.0096 &0.0165 &0.0025 &0.0035 &0.0040 &0.0069\\
 DistMult &0.241 &0.155 &0.263 &0.0105 &0.0139 &0.0171 &0.0278 &0.0036 &0.0046 &0.0055 &0.0086\\
%  ComplEx &0.247 &0.158 &0.275 &0.428 & & & &\\ 
%  RotatE & & & & &  & &\\ 
 ConvE &0.325 &0.237 &0.356 &0.0064 &0.0084 &0.0099 &0.0162 &0.0030 &0.0042 &0.0047 &0.0083\\ 
 R-GCN &0.248 &0.153 &0.258 &-- &-- &-- &-- &-- &-- &-- &--\\
%  \midrule
%  MLN &0.098 &0.067 &0.103 &0.160 &-- &-- &-- &-- &-- &-- &-- &--\\
 pLogicNet &0.332 & 0.237 & 0.367 &-- &-- &-- &-- &-- &-- &-- &--\\
 pGAT &\underline{0.457} & \underline{0.377} & \underline{\textbf{0.494}} &-- &-- &-- &-- &-- &-- &-- &--\\
 \midrule
 BPRMF &-- &-- &-- &0.0274 &0.0348 &0.0428 &0.0658 &0.0086 &0.0109 &0.0129 &0.0200 \\
 NCR &-- &-- &-- &0.0369 &0.0453 &0.0664 &0.0767 &0.0109 &0.0132 &0.0143 &0.0246 \\
 NGCF &-- &-- &-- &\underline{0.0453} &\underline{0.0576} &\underline{0.0715} &\underline{0.1057} &\underline{0.0133} &\underline{0.0173} &\underline{0.0219} &\underline{0.0331} \\
 \midrule
%  $^1\textbf{NCR}_{mod}$ &0.3602 & 0.4217 &0.5085 &0.6985 & \textbf{0.4204*} & \textbf{0.4606*} & \textbf{0.5531} &0.6773 & 0.3256 & 0.3634 & 0.4347 & 0.5514\\
%  \textbf{NLR}(ours) &\textbf{0.3778*} &\textbf{0.4320*} &\textbf{0.5414*} &\textbf{0.7081*} & \textbf{0.4212*} & \textbf{0.4616*} &\textbf{0.5482} & 0.6736 &\textbf{0.3285*} &\textbf{0.3662*} &\textbf{0.4405*} &\textbf{0.5574} \\ 
\textbf{GCR} &\textbf{0.492*} &\textbf{0.490*} &0.493 &\textbf{0.0606*}  &\textbf{0.0829*}  &\textbf{0.0940*}  &\textbf{0.1637*} &\textbf{0.0159*} &\textbf{0.0229*} &\textbf{0.0262*} &\textbf{0.0478*}\\ 
% \textbf{GLN} w/o \textbf{LR} &0.3671 &0.4219 &0.5180 &0.6890 &0.4126 &0.4535 &0.5444 &0.6705 \\
 \midrule
 \midrule
%  $^1$improv. & 16.68\% & 14.44\% & 10.88\% & 7.69\% & 4.94\% & 3.39\% & 1.39\% & - & 5.30\% & 4.67\% & 4.02\% & 2.99\%\\
Improvment &7.66\% &29.97\% & -- &33.77\% &43.92\% &31.47\% &54.87\% &19.55\% &32.37\% &19.63\% &44.41\%\\

% Improvment$^2$ & 4.39\% & 2.14\% & 6.71\% & 2.66\% & 1.53\% & 1.72\% & 1.91\% & 2.26\%\\
 \bottomrule
%  NCR-Reg &0.00 &0.00 &0.00 &0.00 &0.00 &0.00 \\ 
%  \hline
\end{tabular}
\vspace{-5pt}
\end{table*}

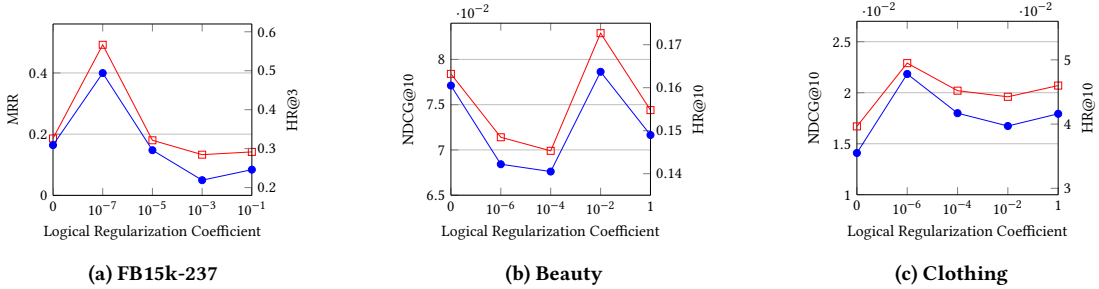
\begin{figure*}
    \begin{subfigure}[b]{0.29\textwidth}
        \centering
        \resizebox{0.8\linewidth}{!}{
\begin{tikzpicture}
\pgfplotsset{
    scale only axis,
    xmin=0, xmax=0.8,
     label style={font=\normalsize}
}

\begin{axis}[
  axis y line*=left,
  ymin=0.00, ymax=0.56,
  xlabel={Logical Regularization Coefficient},
  ylabel={MRR},
  xticklabels={0,0,$10^{-7}$,$10^{-5}$,$10^{-3}$,$10^{-1}$},
  legend pos=south west,
  ymajorgrids=true
]
\addplot[mark=square,red]
  coordinates{
    (0,0.1860)(0.2,0.492)(0.4,0.180)(0.6,0.133)(0.8,0.142)
}; \label{plot_one}
% \addlegendentry{plot 1}
\end{axis}

\begin{axis}[
  axis y line*=right,
  axis x line=none,
  ymin=0.18, ymax=0.62,
  ylabel={HR@3},
%   legend pos=south west,
%   legend style={font=\fontsize{7}{5}\selectfont}
]
\addplot[mark=*,blue]
  coordinates{
    (0,0.309)
    (0.2,0.494)
    (0.4,0.296)
    (0.6,0.219)
    (0.8,0.246)
}; 
% \addlegendimage{/pgfplots/refstyle=plot_one}
% \addlegendentry{HR@5}
% \addlegendentry{NDCG@10}
\end{axis}
\end{tikzpicture}
        }
        \caption{FB15k-237}
        \label{fig:subfig8}
    \end{subfigure}
    % \hskip -8.5ex
\begin{subfigure}[b]{0.30\textwidth}
    \centering
        \resizebox{0.8\linewidth}{!}{
            \begin{tikzpicture}
\pgfplotsset{
    scale only axis,
    xmin=0, xmax=0.8,
    label style={font=\normalsize}
}

\begin{axis}[
  axis y line*=left,
  ymin=0.0650, ymax=0.0840,
  xlabel={Logical Regularization Coefficient},
  ylabel={NDCG@10},
  xticklabels={0,0,$10^{-6}$,$10^{-4}$,$10^{-2}$,1},
  legend pos=south west,
  ymajorgrids=true,
  legend style={font=\fontsize{7}{5}\selectfont}
]
\addplot[mark=square,red]
  coordinates{
  (0,0.0784)(0.2,0.0714)(0.4,0.0699)(0.6,0.0829)(0.8,0.0744)

}; \label{plot_one}
\end{axis}

\begin{axis}[
  axis y line*=right,
  axis x line=none,
  ymin=0.1350, ymax=0.1750,
  ylabel={HR@10},
  xticklabels={0,0,$10^{-4}$,$10^{-3}$,$10^{-2}$,$10^{-1}$,1},
%   legend pos=south east,
]
\addplot[mark=*,blue]
  coordinates{
    (0,0.1605)
    % (0.5,0.6701)
    (0.2,0.1422)
    (0.4,0.1405)
    (0.6,0.1637)
    (0.8,0.1490)
}; 
% \addlegendimage{/pgfplots/refstyle=plot_one}\addlegendentry{HR@10}
% \addlegendentry{NDCG@10}
\end{axis}
\end{tikzpicture}
        }
        \caption{Beauty}   
        \label{fig:subfig9}
    \end{subfigure}
    % \hskip -8.5ex
    \begin{subfigure}[b]{0.285\textwidth}
        \centering
        \resizebox{0.8\linewidth}{!}{
            \begin{tikzpicture}
\pgfplotsset{
    scale only axis,
     label style={font=\normalsize},
    xmin=0, xmax=0.8
}

\begin{axis}[
  axis y line*=left,
  ymin=0.01, ymax=0.027,
  xlabel={Logical Regularization Coefficient},
  ylabel={NDCG@10},
  xticklabels={0,0,$10^{-6}$,$10^{-4}$,$10^{-2}$,1},
%   legend pos=south west,
  ymajorgrids=true,
]
\addplot[mark=square,red]
  coordinates{
    (0,0.0167)(0.2,0.0229)(0.4,0.0202)(0.6,0.0196)(0.8,0.0207)
}; \label{plot_one}
% \addlegendentry{plot 1}
\end{axis}

\begin{axis}[
  axis y line*=right,
  axis x line=none,
  ymin=0.029, ymax=0.056,
  xticklabels={0,0,$10^{-6}$,$10^{-4}$,$10^{-2}$,1},
  ylabel={HR@10},
%   legend pos=south west,
%   legend style={font=\fontsize{7}{5}\selectfont}
]
\addplot[mark=*,blue]
  coordinates{
    (0,0.0355)
    (0.2,0.0478)
    (0.4,0.0417)
    (0.6,0.0397)
    (0.8,0.0416)
}; 
% \addlegendimage{/pgfplots/refstyle=plot_one}\addlegendentry{HR@10}
% \addlegendentry{NDCG@10}
\end{axis}
\end{tikzpicture}
        }
        \caption{Clothing}
        \label{fig:subfig10}
    \end{subfigure}
    % \hskip -11ex
\vspace{-10pt}
\caption{MRR/NDCG@10 (red squared line) and HR@3/HR@10 (blue circled line) on three datasets according to the increment of the logical regularization coefficient $\lambda_r$.}
\vspace{-10pt}
\label{fig:logic_regularizer_weight}
\end{figure*}

\subsection{Overall Performance of GCR (RQ1)}
\label{sec:result_1}
We report the overall performance for graph link prediction and recommendation tasks in Table~\ref{tb:results}. 

For the graph link prediction task, 
% since we use the benchmark dataset FB15k-237 and apply the same settings for the evaluation, we directly extract the baseline results from paper~\cite{qu2019probabilistic}. 
from the results, we see that our GCR model significantly outperforms all the baselines on MRR and Hit@1. The good performance on MRR and Hit@1 indicates that our model can generate high-quality predictions by ranking the correct target at top positions. Although Hit@3 is not better than pGAT, the performance is still competitive. According to the results, we observe that logic-based methods can consistently outperform the other non-logical models. This indicates the effectiveness of applying logic to graph link prediction tasks.

For the recommendation task, our model consistently outperforms all the baselines on all the evaluation metrics. From the reported results, we have the following observations:
\begin{itemize}[leftmargin=*]
    \item Knowledge graph embedding models have relatively worse performance than those recommendation models on the recommendation task. One reason is that the KG embedding models treat each triplet independently while recommendation needs to consider users and items from a collaborative learning perspective. This could limit the KG models to gain a good performance on recommendation tasks. Another reason is that the recommendation data presents more uncertainty than KG data since the recommendation data is recorded from user behaviors while the KG data is mostly fact-based, which is a challenge for the KG embedding methods. 
    \item Among the recommendation baseline models, NGCF outperforms all other baseline methods. This indicates that it is beneficial to incorporate neighborhood information over graphs to make recommendation predictions. 
    \item GCR outperforms NCR. This is because NCR only takes user historical interactions to generate logic expressions. However, GCR not only considers the items that the user interacted with, but also considers which other users interacted with these items. By leveraging the rich information from both user- and item-side, GCR can have a better recommendation quality than NCR.
    \item GCR consistently outperforms all the baselines. In particular, GCR improves over the strongest baseline NGCF on both datasets by at least 19.55\% on NDCG@5. For Hit@10, our model can achieve even 44.41\% improvement on the Clothing dataset. We realize that our model can have higher improvements over baselines when the dataset is more sparse. The Beauty dataset has a density 0.073\% while the Clothing dataset is 0.031\%. This result is reasonable because NGCF needs to aggregate neighborhood information to enhance user and item embedding representations. A very sparse dataset means that the average interactions over each user is limited so that the model cannot aggregate enough neighbor information to promote the representation quality. However, our GCR, by modeling link prediction from logical reasoning perspective, can help to improve the recommendation performance on sparse dataset. We conducted paired $t$-test and the $p$-value < 0.05, which shows that our model has statistical significant improvements over the strongest baseline.
\end{itemize}

\begin{figure*}
%%%%%%%%%%%%% SUB FIG 1 %%%%%%%%%%%%%%%%%%%
\begin{subfigure}[b]{0.249\textwidth}
    \centering
        \resizebox{0.9\linewidth}{!}{
            \begin{tikzpicture}
\pgfplotsset{
    scale only axis,
    label style={font=\normalsize}
}

\begin{axis}[
ybar,
ymin=0, ymax=12800,
bar width=15pt,
enlarge x limits=0.25,
symbolic x coords={<5,<10,<30, >=30},
% nodes near coords=\rotatebox{90}{\scriptsize\pgfmathprintnumber\pgfplotspointmeta},
axis y line*=left,
  xlabel={User Group},
  ylabel={Number of Users},
  xtick=data,
  legend pos=south west,
  ymajorgrids=true,
  legend style={font=\fontsize{7}{5}\selectfont}
]
\addplot[draw=blue,fill=blue!20!white] 
coordinates{
(<5, 6795)(<10, 12123)(<30, 3122)(>=30, 321)}; 
% \addplot[draw=Salmon,fill=Salmon!40!white] coordinates      {(Mar '12,100)      (Apr '12,100)       (May '12,100)}; 
\end{axis}

\begin{axis}[
  axis y line*=right,
  axis x line=none,
  ymin=0.047, ymax=0.184,
  ylabel={Hit@5},
  enlarge x limits=0.25,
  symbolic x coords={<5,<10,<30, >=30},
  xtick=data,
  scaled ticks=false, tick label style={/pgf/number format/fixed},
%   xticklabels={0,0,$10^{-4}$,$10^{-3}$,$10^{-2}$,$10^{-1}$,1},
%   legend pos=south east,
legend style={
                draw=none, % ?
                text depth=0pt,
                at={(0.15,-0.30)},
                anchor=north west,
                legend columns=-1,                        font=\fontsize{8}{5}\selectfont,
                % default spacing:
                column sep=0.1cm,
                % The text "Legend:"
                % /tikz/column 2/.style={column sep=0pt,font=\bfseries},
                %
                % the space between legend image and text:
                /tikz/every odd column/.append style={column sep=0cm},
            }
]
\addplot[mark=*,red, line width=1.5pt]
  coordinates{
    (<5,0.0761)
    (<10,0.0868)
    (<30,0.1045)
    (>=30,0.1572)
}; 
\addlegendentry{GCR}
\addplot[mark=square, blue, line width=1.5pt]
  coordinates{
    (<5,0.0574)
    (<10,0.0577)
    (<30,0.0848)
    (>=30,0.1742)
}; 
\addlegendentry{NGCF} %\addlegendimage{/pgfplots/refstyle=plot_one}\addlegendentry{HR@10}
% \addlegendentry{NDCG@10}
\end{axis}
\end{tikzpicture}
}
        \caption{Beauty Hit@5}   
        \label{fig:subfig9}
    \end{subfigure}
    % \hskip -8.5ex
%%%%%%%%%%%%% SUB FIG 2 %%%%%%%%%%%%%%%%%%%
\begin{subfigure}[b]{0.249\textwidth}
    \centering
        \resizebox{0.9\linewidth}{!}{
            \begin{tikzpicture}
\pgfplotsset{
    scale only axis,
    label style={font=\normalsize}
}

\begin{axis}[
ybar,
ymin=0, ymax=12800,
bar width=15pt,
enlarge x limits=0.25,
symbolic x coords={<5,<10,<30, >=30},
% nodes near coords=\rotatebox{90}{\scriptsize\pgfmathprintnumber\pgfplotspointmeta},
axis y line*=left,
  xlabel={User Group},
  ylabel={Number of Users},
  xtick=data,
  legend pos=south west,
  ymajorgrids=true,
  legend style={font=\fontsize{7}{5}\selectfont}
]
\addplot[draw=blue,fill=blue!20!white] 
coordinates{
(<5, 6795)(<10, 12123)(<30, 3122)(>=30, 321)}; 
% \addplot[draw=Salmon,fill=Salmon!40!white] coordinates      {(Mar '12,100)      (Apr '12,100)       (May '12,100)}; 
\end{axis}

\begin{axis}[
  axis y line*=right,
  axis x line=none,
  ymin=0.070, ymax=0.28,
  ylabel={Hit@10},
  enlarge x limits=0.25,
  symbolic x coords={<5,<10,<30, >=30},
  xtick=data,
%   xticklabels={0,0,$10^{-4}$,$10^{-3}$,$10^{-2}$,$10^{-1}$,1},
%   legend pos=south east,
legend style={
                draw=none, % ?
                text depth=0pt,
                at={(0.15,-0.30)},
                anchor=north west,
                legend columns=-1,                        font=\fontsize{8}{5}\selectfont,
                % default spacing:
                column sep=0.1cm,
                % The text "Legend:"
                % /tikz/column 2/.style={column sep=0pt,font=\bfseries},
                %
                % the space between legend image and text:
                /tikz/every odd column/.append style={column sep=0cm},
            }
]
]
\addplot[mark=*,red, line width=1.5pt]
  coordinates{
    (<5,0.1292)
    (<10,0.1489)
    (<30,0.1845)
    (>=30,0.2452)
}; 
\addlegendentry{GCR}
\addplot[mark=square, blue, line width=1.5pt]
  coordinates{
    (<5,0.0849)
    (<10,0.0897)
    (<30,0.1281)
    (>=30,0.2677)
}; 
\addlegendentry{NGCF}
% \addlegendimage{/pgfplots/refstyle=plot_one}\addlegendentry{HR@10}
% \addlegendentry{NDCG@10}
\end{axis}
\end{tikzpicture}
}
        \caption{Beauty Hit@10}   
        \label{fig:subfig9}
\end{subfigure}
%%%%%%%%%%%%% SUB FIG 3 %%%%%%%%%%%%%%%%%%%
\begin{subfigure}[b]{0.242\textwidth}
    \centering
        \resizebox{0.9\linewidth}{!}{
            \begin{tikzpicture}
\pgfplotsset{
    scale only axis,
    label style={font=\normalsize}
}

\begin{axis}[
ybar,
ymin=0, ymax=16990,
bar width=15pt,
enlarge x limits=0.25,
symbolic x coords={<5,<7,<15, >=15},
% nodes near coords=\rotatebox{90}{\scriptsize\pgfmathprintnumber\pgfplotspointmeta},
axis y line*=left,
  xlabel={User Group},
  ylabel={Number of Users},
  xtick=data,
  legend pos=south west,
  ymajorgrids=true,
  legend style={font=\fontsize{7}{5}\selectfont}
]
]
\addplot[draw=blue,fill=blue!20!white] 
coordinates{
(<5, 14381)(<7, 15994)(<15, 8258)(>=15, 748)}; 
% \addplot[draw=Salmon,fill=Salmon!40!white] coordinates      {(Mar '12,100)      (Apr '12,100)       (May '12,100)}; 
\end{axis}

\begin{axis}[
  axis y line*=right,
  axis x line=none,
  ymin=0.014, ymax=0.038,
  ylabel={Hit@5},
  enlarge x limits=0.25,
  symbolic x coords={<5,<7,<15, >=15},
  xtick=data,
  scaled ticks=false, tick label style={/pgf/number format/fixed},
%   xticklabels={0,0,$10^{-4}$,$10^{-3}$,$10^{-2}$,$10^{-1}$,1},
%   legend pos=south east,
legend style={
                draw=none, % ?
                text depth=0pt,
                at={(0.15,-0.30)},
                anchor=north west,
                legend columns=-1,                        font=\fontsize{8}{5}\selectfont,
                % default spacing:
                column sep=0.1cm,
                % The text "Legend:"
                % /tikz/column 2/.style={column sep=0pt,font=\bfseries},
                %
                % the space between legend image and text:
                /tikz/every odd column/.append style={column sep=0cm},
            }
]
]
\addplot[mark=*,red, line width=1.5pt]
  coordinates{
    (<5,0.0299)
    (<7,0.0245)
    (<15,0.0270)
    (>=15,0.0260)
}; 
\addlegendentry{GCR}
\addplot[mark=square, blue, line width=1.5pt]
  coordinates{
    (<5,0.0188)
    (<7,0.0184)
    (<15,0.0234)
    (>=15,0.0273)
}; 
\addlegendentry{NGCF}
% \addlegendimage{/pgfplots/refstyle=plot_one}\addlegendentry{HR@10}
% \addlegendentry{NDCG@10}
\end{axis}
\end{tikzpicture}
}
        \caption{Clothing Hit@5}   
        \label{fig:subfig9}
\end{subfigure}
%%%%%%%%%%%%% SUB FIG 4 %%%%%%%%%%%%%%%%%%%
\begin{subfigure}[b]{0.242\textwidth}
    \centering
        \resizebox{0.9\linewidth}{!}{
            \begin{tikzpicture}
\pgfplotsset{
    scale only axis,
    label style={font=\normalsize}
}

\begin{axis}[
ybar,
ymin=0, ymax=16990,
bar width=15pt,
enlarge x limits=0.25,
symbolic x coords={<5,<7,<15, >=15},
% nodes near coords=\rotatebox{90}{\scriptsize\pgfmathprintnumber\pgfplotspointmeta},
axis y line*=left,
  xlabel={User Group},
  ylabel={Number of Users},
  xtick=data,
  legend pos=south west,
  ymajorgrids=true,
  legend style={font=\fontsize{7}{5}\selectfont}
]
\addplot[draw=blue,fill=blue!20!white] 
coordinates{
(<5, 14381)(<7, 15994)(<15, 8250)(>=15, 748)}; 
% \addplot[draw=Salmon,fill=Salmon!40!white] coordinates      {(Mar '12,100)      (Apr '12,100)       (May '12,100)}; 
\end{axis}

\begin{axis}[
  axis y line*=right,
  axis x line=none,
  ymin=0.020, ymax=0.062,
  ylabel={Hit@10},
  enlarge x limits=0.25,
  symbolic x coords={<5,<7,<15, >=15},
  xtick=data,
  scaled ticks=false, tick label style={/pgf/number format/fixed},
%   xticklabels={0,0,$10^{-4}$,$10^{-3}$,$10^{-2}$,$10^{-1}$,1},
%   legend pos=south east,
legend style={
                draw=none, % ?
                text depth=0pt,
                at={(0.15,-0.30)},
                anchor=north west,
                legend columns=-1,                        font=\fontsize{8}{5}\selectfont,
                % default spacing:
                column sep=0.1cm,
                % The text "Legend:"
                % /tikz/column 2/.style={column sep=0pt,font=\bfseries},
                %
                % the space between legend image and text:
                /tikz/every odd column/.append style={column sep=0cm},
            }
]
]
\addplot[mark=*,red, line width=1.5pt]
  coordinates{
    (<5,0.0533)
    (<7,0.0453)
    (<15,0.0477)
    (>=15,0.0526)
}; 
\addlegendentry{GCR}
\addplot[mark=square, blue, line width=1.5pt]
  coordinates{
    (<5,0.0308)
    (<7,0.0301)
    (<15,0.0364)
    (>=15,0.0594)
}; 
\addlegendentry{NGCF}
% \addlegendimage{/pgfplots/refstyle=plot_one}\addlegendentry{HR@10}
% \addlegendentry{NDCG@10}
\end{axis}
\end{tikzpicture}
}
        \caption{Clothing Hit@10}   
        \label{fig:subfig9}
\end{subfigure}
    % \hskip -11ex
\vspace{-10pt}
\caption{Performance comparision between GCR and NGCF on Beauty and Clothing datasets. The histograms represent the total number of users in each group, the lines indicate the performance trend with the growing number of per user interactions.
% $w.r.t$ Hit@5 and Hit@10.
}
\vspace{-10pt}
\label{fig:sparsity}
\end{figure*}
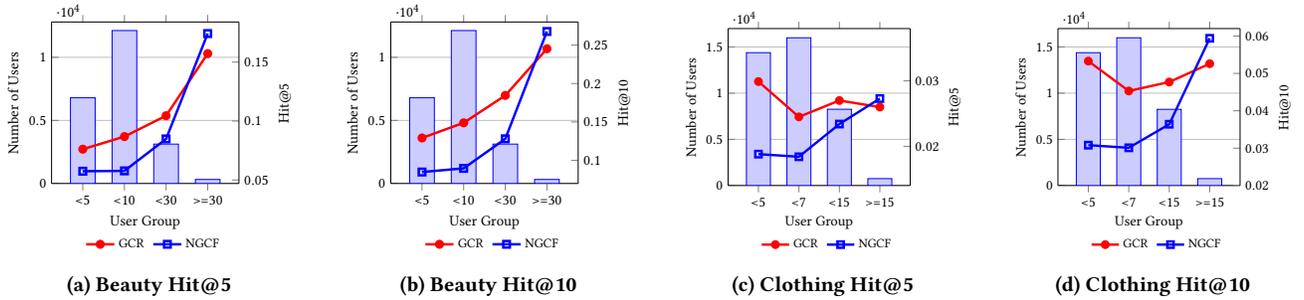

\vspace{-2ex}
\subsection{Impact of Logical Regularization (RQ2)}
\label{sec:result_2}
In this section, we answer the question that if the logical regularization helps the learning process. We conduct experiments by tuning the logical regularization coefficient $\lambda_l$ in $[0, 10^{-7}, 10^{-5}, 10^{-3}, 10^{-1}]$ for \textit{FB15k-237} and $[0, 10^{-6}, 10^{-4}, 10^{-2}, 1]$ for \textit{Beauty} and \textit{Clothing}. We show how performance changes w.r.t MRR, Hit Rate and NDCG in Figure~\ref{fig:logic_regularizer_weight}. We have two major observations from the results:

\begin{itemize}[leftmargin=*]
    \item The results show that logical regularization do help to improve the performance when comparing the results of non-logic model ($\lambda_{l}=0$) and logic-regularized models ($\lambda_{l}\neq0$). However, how strong the regularization should be added to the neural network need to be carefully adjusted, similar to the observations in \cite{chen2020neural}.
    \item Sparser data needs a relatively smaller logical regularization coefficient. For the Beauty and Clothing datasets, which are bipartite graphs, their densities are 0.073\% and 0.031\%, respectively. For FB15k-237, which is a multi-relational graph, the density is $\frac{|\mathcal{T}|}{|\mathcal{V}|\times |\mathcal{V} - 1| \times |\mathcal{R}|}\times 100\% \approx 0.0006\%$. This is because we not only need to decide if an entity pair will be connected but also need to decide the type of relation between them, which is different from the recommendation bipartite graphs. For the most sparse data FB15k-237, the best logic regularization weight is $10^{-7}$, while the best weight for the most dense dataset among the three is $10^{-2}$. The reason for the observation is that there is a trade-off between the prediction loss and the logical loss. The model needs to learn useful information from limited data to generate good predictions. For the sparse FB15k-237 dataset, the model is very sensitive to large logical regularization weights because the logical loss will dominate the total loss when training data is insufficient for the prediction loss. However, for Clothing dataset, which is about 50 times denser than FB15k-237, we see that
    % they both show similar pattern in the given figure. However, 
    the model is not that sensitive to large logical regularization weights. Even with a higher regularization weight, the model still achieves better performance than non-logic model that $\lambda_l=0$. 
\end{itemize}

% Apart from the insights we gain from the analysis, we think one reason for the logical regularization could help to improve the performance is that such constraint could help the model be optimized towards a correct direction. Since logical information is naturally existed inside most of real-world data in either hard or soft form, the logical regularization can help the model tend to the optimal solution during the optimization process. 

% \vspace{-5pt}
\subsection{Impact of Sparsity Levels (RQ3)}
\label{sec:result_3}
The sparsity issue brought by data incompleteness may limit the embedding quality of prediction models. When the data is insufficient, it is difficult for models to capture the relations between entity pairs, and thus influence the quality of the generated predictions. This issue would especially affect the link prediction models since they usually relies on collective information for model learning.
% similarity-based algorithms such as collaborative filtering or translation-based methods. 
In this section, we explore whether logical reasoning models can help to improve the prediction performance when the data is sparse. With this consideration, we conduct an experiment by evaluating the model performance over different data groups that have different sparsity. For better visualization of the results, we perform the experiments on the two bipartite graphs.

In particular, we split the users in the testing set into different groups based on their total number of interactions in the training data. Take the Beauty dataset as an example, users are divided into four groups, corresponding to the users whose number of interactions is in $[1,5)$, $[5,10)$, $[10,30)$ and $[30,\infty)$, respectively. We compare our model with the strong baseline NGCF and report the results with respect to Hit@5 and Hit@10 in Figure~\ref{fig:sparsity}. Since similar trend is also observed on the NDCG metric, we do not plot the NDCG results to keep the figure clarity. 

From the experiments, we see that our GCR model has significantly better performance than NGCF on sparse user groups. When the user has more interactions, the performance of NGCF can be better than ours. This observation can be explained by the underlying modeling mechanism of NGCG and GCR. NGCF needs to take the neighborhood information to enrich the node embeddings. For the users with very few interactions, it would be challenging for NGCF to capture the user similarities. Although the GCR model also relies on the neighborhood information, it benefits from two special advantages. First, the model can leverage both neighbour node and neighbour link information, and second, the logic component helps to model the logical relationship among the limited neighbourhood entities rather than merely relying on the associative node similarity information for prediction.
The good performance on sparse user groups show that our logical reasoning-based model helps to improve the recommendation quality on sparse data. This is an important advantage of our model, since users with fewer interactions are the majority, as shown in Figure~\ref{fig:sparsity}. 
% The prediction quality on sparse user groups will significantly affect the overall performance. 

\vspace{-2ex}
\section{Conclusions and Future Work}
\label{sec:conclusion}
% 1-hop -> multi-hop
In this paper, we propose to model link prediction as a reasoning problem over graphs. Specifically, we propose a Graph Collaborative Reasoning (GCR) approach, which takes the neighborhood link information to predict the connections in a latent reasoning space. Experiments on two representative link prediction tasks---graph link prediction and recommendation---show the effectiveness of the model, especially for link prediction on sparse data.
% Specifically, we model the sub-graph, which contains all the necessary neighbor links for a prediction task, as a logical expression and convert such link prediction to a logical satisfiability problem. Then the answer of the validity of an unknown triplet becomes the evaluation result of the corresponding logical expression. We grant Graph Logic Network as the name of our model. This name represents not only for its ability of using logical reasoning, but also for its ability of aggregating neighbor triplets information for link prediction task, which is similar to the mechanism of GNN. 

We believe enabling the ability of reasoning over graphs is important for future cognitive intelligent systems. This work is just one of our first steps towards this goal, and there is still much room for future improvements. In this paper, we only used the one-hop neighborhood links, while in the future we will extend to multi-hop reasoning over graphs based on the GCR framework to model hierarchical data structure. Besides the knowledge graph and recommendation tasks considered in this work, graph collaborative reasoning may also help other intelligent tasks such as question answering, molecular graph modeling, entity search and conversational systems, which we will explore in the future.
% In the future, we will explore the way of including multi-hop relations into logical reasoning process for link prediction.

% \begin{acks}
% We thank the reviewers for the reviews and suggestions. 
% This work was supported in part by NSF IIS-1910154, IIS-2007907, IIS-2046457 and CCF-2124155. Any opinions and findings in this material are those of the authors and do not necessarily reflect those of the sponsors.
% \end{acks}

\bibliographystyle{ACM-Reference-Format}
\balance
\bibliography{sample-base}


\begin{thebibliography}{45}

%%% the Web address.

\ifx \showCODEN    \undefined \def \showCODEN     #1{\unskip}     \fi
\ifx \showDOI      \undefined \def \showDOI       #1{#1}\fi
\ifx \showISBNx    \undefined \def \showISBNx     #1{\unskip}     \fi
\ifx \showISBNxiii \undefined \def \showISBNxiii  #1{\unskip}     \fi
\ifx \showISSN     \undefined \def \showISSN      #1{\unskip}     \fi
\ifx \showLCCN     \undefined \def \showLCCN      #1{\unskip}     \fi
\ifx \shownote     \undefined \def \shownote      #1{#1}          \fi
\ifx \showarticletitle \undefined \def \showarticletitle #1{#1}   \fi
\ifx \showURL      \undefined \def \showURL       {\relax}        \fi
% invisible to TeX
\providecommand\bibfield[2]{#2}
\providecommand\bibinfo[2]{#2}
\providecommand\natexlab[1]{#1}
\providecommand\showeprint[2][]{arXiv:#2}

\bibitem[\protect\citeauthoryear{Arora}{Arora}{2020}]%
        {arora2020survey}
\bibfield{author}{\bibinfo{person}{Siddhant Arora}.}
  \bibinfo{year}{2020}\natexlab{}.
\newblock \showarticletitle{A Survey on Graph Neural Networks for Knowledge
  Graph Completion}.
\newblock \bibinfo{journal}{\emph{arXiv preprint arXiv:2007.12374}}
  (\bibinfo{year}{2020}).
\newblock


\bibitem[\protect\citeauthoryear{Bordes, Usunier, Garcia-Duran, Weston, and
  Yakhnenko}{Bordes et~al\mbox{.}}{2013}]%
        {bordes2013translating}
\bibfield{author}{\bibinfo{person}{Antoine Bordes}, \bibinfo{person}{Nicolas
  Usunier}, \bibinfo{person}{Alberto Garcia-Duran}, \bibinfo{person}{Jason
  Weston}, {and} \bibinfo{person}{Oksana Yakhnenko}.}
  \bibinfo{year}{2013}\natexlab{}.
\newblock \showarticletitle{Translating embeddings for modeling
  multi-relational data}. In \bibinfo{booktitle}{\emph{Advances in neural
  information processing systems}}. \bibinfo{pages}{2787--2795}.
\newblock


\bibitem[\protect\citeauthoryear{Chen, Shi, Li, and Zhang}{Chen
  et~al\mbox{.}}{2021}]%
        {chen2020neural}
\bibfield{author}{\bibinfo{person}{Hanxiong Chen}, \bibinfo{person}{Shaoyun
  Shi}, \bibinfo{person}{Yunqi Li}, {and} \bibinfo{person}{Yongfeng Zhang}.}
  \bibinfo{year}{2021}\natexlab{}.
\newblock \showarticletitle{Neural Collaborative Reasoning}. In
  \bibinfo{booktitle}{\emph{Proceedings of the 30th Web Conference (WWW)}}.
\newblock


\bibitem[\protect\citeauthoryear{Demeester, Rockt{\"a}schel, and
  Riedel}{Demeester et~al\mbox{.}}{2016}]%
        {demeester2016lifted}
\bibfield{author}{\bibinfo{person}{Thomas Demeester}, \bibinfo{person}{Tim
  Rockt{\"a}schel}, {and} \bibinfo{person}{Sebastian Riedel}.}
  \bibinfo{year}{2016}\natexlab{}.
\newblock \showarticletitle{Lifted Rule Injection for Relation Embeddings}. In
  \bibinfo{booktitle}{\emph{Proceedings of the 2016 Conference on Empirical
  Methods in Natural Language Processing}}. \bibinfo{pages}{1389--1399}.
\newblock


\bibitem[\protect\citeauthoryear{Dettmers, Minervini, Stenetorp, and
  Riedel}{Dettmers et~al\mbox{.}}{2018}]%
        {dettmers2018convolutional}
\bibfield{author}{\bibinfo{person}{T Dettmers}, \bibinfo{person}{P Minervini},
  \bibinfo{person}{P Stenetorp}, {and} \bibinfo{person}{S Riedel}.}
  \bibinfo{year}{2018}\natexlab{}.
\newblock \showarticletitle{Convolutional 2D knowledge graph embeddings}. In
  \bibinfo{booktitle}{\emph{32nd AAAI Conference on Artificial Intelligence,
  AAAI 2018}}, Vol.~\bibinfo{volume}{32}. AAI Publications,
  \bibinfo{pages}{1811--1818}.
\newblock


\bibitem[\protect\citeauthoryear{Ding, Wang, Wang, and Guo}{Ding
  et~al\mbox{.}}{2018}]%
        {ding2018improving}
\bibfield{author}{\bibinfo{person}{Boyang Ding}, \bibinfo{person}{Quan Wang},
  \bibinfo{person}{Bin Wang}, {and} \bibinfo{person}{Li Guo}.}
  \bibinfo{year}{2018}\natexlab{}.
\newblock \showarticletitle{Improving Knowledge Graph Embedding Using Simple
  Constraints}. In \bibinfo{booktitle}{\emph{Proceedings of the 56th Annual
  Meeting of the Association for Computational Linguistics (Volume 1: Long
  Papers)}}. \bibinfo{pages}{110--121}.
\newblock


\bibitem[\protect\citeauthoryear{Gilmer, Schoenholz, Riley, Vinyals, and
  Dahl}{Gilmer et~al\mbox{.}}{2017}]%
        {gilmer2017neural}
\bibfield{author}{\bibinfo{person}{Justin Gilmer}, \bibinfo{person}{Samuel~S
  Schoenholz}, \bibinfo{person}{Patrick~F Riley}, \bibinfo{person}{Oriol
  Vinyals}, {and} \bibinfo{person}{George~E Dahl}.}
  \bibinfo{year}{2017}\natexlab{}.
\newblock \showarticletitle{Neural Message Passing for Quantum Chemistry}. In
  \bibinfo{booktitle}{\emph{ICML}}.
\newblock


\bibitem[\protect\citeauthoryear{Guo, Li, Hui, Meng, Ma, Liu, Wang, Zhai, and
  Zhang}{Guo et~al\mbox{.}}{2020}]%
        {guo2020knowledge}
\bibfield{author}{\bibinfo{person}{Shu Guo}, \bibinfo{person}{Lin Li},
  \bibinfo{person}{Zhen Hui}, \bibinfo{person}{Lingshuai Meng},
  \bibinfo{person}{Bingnan Ma}, \bibinfo{person}{Wei Liu},
  \bibinfo{person}{Lihong Wang}, \bibinfo{person}{Haibin Zhai}, {and}
  \bibinfo{person}{Hong Zhang}.} \bibinfo{year}{2020}\natexlab{}.
\newblock \showarticletitle{Knowledge Graph Embedding Preserving Soft Logical
  Regularity}. In \bibinfo{booktitle}{\emph{Proceedings of the 29th ACM
  International Conference on Information \& Knowledge Management}}.
  \bibinfo{pages}{425--434}.
\newblock


\bibitem[\protect\citeauthoryear{Guo, Wang, Wang, Wang, and Guo}{Guo
  et~al\mbox{.}}{2016}]%
        {guo2016jointly}
\bibfield{author}{\bibinfo{person}{Shu Guo}, \bibinfo{person}{Quan Wang},
  \bibinfo{person}{Lihong Wang}, \bibinfo{person}{Bin Wang}, {and}
  \bibinfo{person}{Li Guo}.} \bibinfo{year}{2016}\natexlab{}.
\newblock \showarticletitle{Jointly embedding knowledge graphs and logical
  rules}. In \bibinfo{booktitle}{\emph{Proceedings of the 2016 Conference on
  Empirical Methods in Natural Language Processing}}.
  \bibinfo{pages}{192--202}.
\newblock


\bibitem[\protect\citeauthoryear{Guo, Wang, Wang, Wang, and Guo}{Guo
  et~al\mbox{.}}{2018}]%
        {guo2018knowledge}
\bibfield{author}{\bibinfo{person}{Shu Guo}, \bibinfo{person}{Quan Wang},
  \bibinfo{person}{Lihong Wang}, \bibinfo{person}{Bin Wang}, {and}
  \bibinfo{person}{Li Guo}.} \bibinfo{year}{2018}\natexlab{}.
\newblock \showarticletitle{Knowledge graph embedding with iterative guidance
  from soft rules}.
\newblock \bibinfo{journal}{\emph{AAAI}} (\bibinfo{year}{2018}).
\newblock


\bibitem[\protect\citeauthoryear{Hamilton, Ying, and Leskovec}{Hamilton
  et~al\mbox{.}}{2017}]%
        {hamilton2017inductive}
\bibfield{author}{\bibinfo{person}{Will Hamilton}, \bibinfo{person}{Zhitao
  Ying}, {and} \bibinfo{person}{Jure Leskovec}.}
  \bibinfo{year}{2017}\natexlab{}.
\newblock \showarticletitle{Inductive representation learning on large graphs}.
  In \bibinfo{booktitle}{\emph{Advances in neural information processing
  systems}}. \bibinfo{pages}{1024--1034}.
\newblock


\bibitem[\protect\citeauthoryear{Harsha~Vardhan, Jia, and Kok}{Harsha~Vardhan
  et~al\mbox{.}}{2020}]%
        {harsha2020probabilistic}
\bibfield{author}{\bibinfo{person}{L~Vivek Harsha~Vardhan},
  \bibinfo{person}{Guo Jia}, {and} \bibinfo{person}{Stanley Kok}.}
  \bibinfo{year}{2020}\natexlab{}.
\newblock \showarticletitle{Probabilistic Logic Graph Attention Networks for
  Reasoning}. In \bibinfo{booktitle}{\emph{Companion Proceedings of the Web
  Conference 2020}}. \bibinfo{pages}{669--673}.
\newblock


\bibitem[\protect\citeauthoryear{Ji, He, Xu, Liu, and Zhao}{Ji
  et~al\mbox{.}}{2015}]%
        {ji2015knowledge}
\bibfield{author}{\bibinfo{person}{Guoliang Ji}, \bibinfo{person}{Shizhu He},
  \bibinfo{person}{Liheng Xu}, \bibinfo{person}{Kang Liu}, {and}
  \bibinfo{person}{Jun Zhao}.} \bibinfo{year}{2015}\natexlab{}.
\newblock \showarticletitle{Knowledge graph embedding via dynamic mapping
  matrix}. In \bibinfo{booktitle}{\emph{Proceedings of the 53rd annual meeting
  of the association for computational linguistics and the 7th international
  joint conference on natural language processing (volume 1: Long papers)}}.
  \bibinfo{pages}{687--696}.
\newblock


\bibitem[\protect\citeauthoryear{Kingma and Ba}{Kingma and Ba}{2014}]%
        {kingma2014adam}
\bibfield{author}{\bibinfo{person}{Diederik~P Kingma} {and}
  \bibinfo{person}{Jimmy Ba}.} \bibinfo{year}{2014}\natexlab{}.
\newblock \showarticletitle{Adam: A method for stochastic optimization}.
\newblock \bibinfo{journal}{\emph{arXiv preprint arXiv:1412.6980}}
  (\bibinfo{year}{2014}).
\newblock


\bibitem[\protect\citeauthoryear{Kipf and Welling}{Kipf and Welling}{2017}]%
        {Kipf:2016tc}
\bibfield{author}{\bibinfo{person}{Thomas~N. Kipf} {and} \bibinfo{person}{Max
  Welling}.} \bibinfo{year}{2017}\natexlab{}.
\newblock \showarticletitle{{Semi-Supervised Classification with Graph
  Convolutional Networks}}. In \bibinfo{booktitle}{\emph{Proceedings of the 5th
  International Conference on Learning Representations}}
  \emph{(\bibinfo{series}{ICLR '17})}.
\newblock


\bibitem[\protect\citeauthoryear{Koren, Bell, and Volinsky}{Koren
  et~al\mbox{.}}{2009}]%
        {koren2009matrix}
\bibfield{author}{\bibinfo{person}{Yehuda Koren}, \bibinfo{person}{Robert
  Bell}, {and} \bibinfo{person}{Chris Volinsky}.}
  \bibinfo{year}{2009}\natexlab{}.
\newblock \showarticletitle{Matrix factorization techniques for recommender
  systems}.
\newblock \bibinfo{journal}{\emph{Computer}} \bibinfo{number}{8}
  (\bibinfo{year}{2009}), \bibinfo{pages}{30--37}.
\newblock


\bibitem[\protect\citeauthoryear{Lin, Liu, Sun, Liu, and Zhu}{Lin
  et~al\mbox{.}}{2015}]%
        {lin2015learning}
\bibfield{author}{\bibinfo{person}{Yankai Lin}, \bibinfo{person}{Zhiyuan Liu},
  \bibinfo{person}{Maosong Sun}, \bibinfo{person}{Yang Liu}, {and}
  \bibinfo{person}{Xuan Zhu}.} \bibinfo{year}{2015}\natexlab{}.
\newblock \showarticletitle{Learning entity and relation embeddings for
  knowledge graph completion}. In \bibinfo{booktitle}{\emph{Proceedings of the
  AAAI Conference on Artificial Intelligence}}, Vol.~\bibinfo{volume}{29}.
\newblock


\bibitem[\protect\citeauthoryear{McAuley, Targett, Shi, and van~den
  Hengel}{McAuley et~al\mbox{.}}{2015}]%
        {mcauley2015image}
\bibfield{author}{\bibinfo{person}{Julian McAuley},
  \bibinfo{person}{Christopher Targett}, \bibinfo{person}{Qinfeng Shi}, {and}
  \bibinfo{person}{Anton van~den Hengel}.} \bibinfo{year}{2015}\natexlab{}.
\newblock \showarticletitle{Image-based recommendations on styles and
  substitutes}. In \bibinfo{booktitle}{\emph{SIGIR}}. ACM.
\newblock


\bibitem[\protect\citeauthoryear{Minervini, Costabello, Mu{\~n}oz,
  Nov{\'a}{\v{c}}ek, and Vandenbussche}{Minervini et~al\mbox{.}}{2017}]%
        {minervini2017regularizing}
\bibfield{author}{\bibinfo{person}{Pasquale Minervini}, \bibinfo{person}{Luca
  Costabello}, \bibinfo{person}{Emir Mu{\~n}oz}, \bibinfo{person}{V{\'\i}t
  Nov{\'a}{\v{c}}ek}, {and} \bibinfo{person}{Pierre-Yves Vandenbussche}.}
  \bibinfo{year}{2017}\natexlab{}.
\newblock \showarticletitle{Regularizing knowledge graph embeddings via
  equivalence and inversion axioms}. In \bibinfo{booktitle}{\emph{Joint
  European Conference on Machine Learning and Knowledge Discovery in
  Databases}}. Springer, \bibinfo{pages}{668--683}.
\newblock


\bibitem[\protect\citeauthoryear{Nathani, Chauhan, Sharma, and Kaul}{Nathani
  et~al\mbox{.}}{2019}]%
        {nathani2019learning}
\bibfield{author}{\bibinfo{person}{Deepak Nathani}, \bibinfo{person}{Jatin
  Chauhan}, \bibinfo{person}{Charu Sharma}, {and} \bibinfo{person}{Manohar
  Kaul}.} \bibinfo{year}{2019}\natexlab{}.
\newblock \showarticletitle{Learning Attention-based Embeddings for Relation
  Prediction in Knowledge Graphs}. In \bibinfo{booktitle}{\emph{Proceedings of
  the 57th Annual Meeting of the Association for Computational Linguistics}}.
  \bibinfo{pages}{4710--4723}.
\newblock


\bibitem[\protect\citeauthoryear{Nguyen, Nguyen, Phung, et~al\mbox{.}}{Nguyen
  et~al\mbox{.}}{2018}]%
        {nguyen2018novel}
\bibfield{author}{\bibinfo{person}{Tu~Dinh Nguyen}, \bibinfo{person}{Dat~Quoc
  Nguyen}, \bibinfo{person}{Dinh Phung}, {et~al\mbox{.}}}
  \bibinfo{year}{2018}\natexlab{}.
\newblock \showarticletitle{A Novel Embedding Model for Knowledge Base
  Completion Based on Convolutional Neural Network}. In
  \bibinfo{booktitle}{\emph{Proceedings of the 2018 Conference of the North
  American Chapter of the Association for Computational Linguistics: Human
  Language Technologies, Volume 2 (Short Papers)}}. \bibinfo{pages}{327--333}.
\newblock


\bibitem[\protect\citeauthoryear{Nickel, Rosasco, and Poggio}{Nickel
  et~al\mbox{.}}{2016}]%
        {nickel2016holographic}
\bibfield{author}{\bibinfo{person}{Maximilian Nickel}, \bibinfo{person}{Lorenzo
  Rosasco}, {and} \bibinfo{person}{Tomaso Poggio}.}
  \bibinfo{year}{2016}\natexlab{}.
\newblock \showarticletitle{Holographic embeddings of knowledge graphs}. In
  \bibinfo{booktitle}{\emph{Proceedings of the AAAI Conference on Artificial
  Intelligence}}, Vol.~\bibinfo{volume}{30}.
\newblock


\bibitem[\protect\citeauthoryear{Nickel, Tresp, and Kriegel}{Nickel
  et~al\mbox{.}}{2011}]%
        {nickel2011three}
\bibfield{author}{\bibinfo{person}{Maximilian Nickel}, \bibinfo{person}{Volker
  Tresp}, {and} \bibinfo{person}{Hans-Peter Kriegel}.}
  \bibinfo{year}{2011}\natexlab{}.
\newblock \showarticletitle{A three-way model for collective learning on
  multi-relational data}. In \bibinfo{booktitle}{\emph{Icml}}.
\newblock


\bibitem[\protect\citeauthoryear{Qu and Tang}{Qu and Tang}{2019}]%
        {qu2019probabilistic}
\bibfield{author}{\bibinfo{person}{Meng Qu} {and} \bibinfo{person}{Jian Tang}.}
  \bibinfo{year}{2019}\natexlab{}.
\newblock \showarticletitle{Probabilistic logic neural networks for reasoning}.
\newblock \bibinfo{journal}{\emph{Advances in neural information processing
  systems}}  \bibinfo{volume}{32} (\bibinfo{year}{2019}),
  \bibinfo{pages}{7712--7722}.
\newblock


\bibitem[\protect\citeauthoryear{Ren and Leskovec}{Ren and Leskovec}{2020}]%
        {ren2020beta}
\bibfield{author}{\bibinfo{person}{Hongyu Ren} {and} \bibinfo{person}{Jure
  Leskovec}.} \bibinfo{year}{2020}\natexlab{}.
\newblock \showarticletitle{Beta Embeddings for Multi-Hop Logical Reasoning in
  Knowledge Graphs}.
\newblock \bibinfo{journal}{\emph{arXiv preprint arXiv:2010.11465}}
  (\bibinfo{year}{2020}).
\newblock


\bibitem[\protect\citeauthoryear{Rendle, Freudenthaler, Gantner, and
  Schmidt-Thieme}{Rendle et~al\mbox{.}}{2009}]%
        {bpr}
\bibfield{author}{\bibinfo{person}{Steffen Rendle}, \bibinfo{person}{Christoph
  Freudenthaler}, \bibinfo{person}{Zeno Gantner}, {and} \bibinfo{person}{Lars
  Schmidt-Thieme}.} \bibinfo{year}{2009}\natexlab{}.
\newblock \showarticletitle{BPR: Bayesian personalized ranking from implicit
  feedback}. In \bibinfo{booktitle}{\emph{Proceedings of the 25th conference on
  uncertainty in artificial intelligence}}. AUAI Press,
  \bibinfo{pages}{452--461}.
\newblock


\bibitem[\protect\citeauthoryear{Rockt{\"a}schel, Singh, and
  Riedel}{Rockt{\"a}schel et~al\mbox{.}}{2015}]%
        {rocktaschel2015injecting}
\bibfield{author}{\bibinfo{person}{Tim Rockt{\"a}schel},
  \bibinfo{person}{Sameer Singh}, {and} \bibinfo{person}{Sebastian Riedel}.}
  \bibinfo{year}{2015}\natexlab{}.
\newblock \showarticletitle{Injecting logical background knowledge into
  embeddings for relation extraction}. In \bibinfo{booktitle}{\emph{Proceedings
  of the 2015 Conference of the North American Chapter of the Association for
  Computational Linguistics: Human Language Technologies}}.
  \bibinfo{pages}{1119--1129}.
\newblock


\bibitem[\protect\citeauthoryear{Rossi, Firmani, Matinata, Merialdo, and
  Barbosa}{Rossi et~al\mbox{.}}{2020}]%
        {rossi2020knowledge}
\bibfield{author}{\bibinfo{person}{Andrea Rossi}, \bibinfo{person}{Donatella
  Firmani}, \bibinfo{person}{Antonio Matinata}, \bibinfo{person}{Paolo
  Merialdo}, {and} \bibinfo{person}{Denilson Barbosa}.}
  \bibinfo{year}{2020}\natexlab{}.
\newblock \showarticletitle{Knowledge Graph Embedding for Link Prediction: A
  Comparative Analysis}.
\newblock \bibinfo{journal}{\emph{arXiv preprint arXiv:2002.00819}}
  (\bibinfo{year}{2020}).
\newblock


\bibitem[\protect\citeauthoryear{Rumelhart, Hinton, and Williams}{Rumelhart
  et~al\mbox{.}}{1986}]%
        {rumelhart1986learning}
\bibfield{author}{\bibinfo{person}{David~E Rumelhart},
  \bibinfo{person}{Geoffrey~E Hinton}, {and} \bibinfo{person}{Ronald~J
  Williams}.} \bibinfo{year}{1986}\natexlab{}.
\newblock \showarticletitle{Learning representations by back-propagating
  errors}.
\newblock \bibinfo{journal}{\emph{nature}} \bibinfo{volume}{323},
  \bibinfo{number}{6088} (\bibinfo{year}{1986}), \bibinfo{pages}{533--536}.
\newblock


\bibitem[\protect\citeauthoryear{Schlichtkrull, Kipf, Bloem, Van Den~Berg,
  Titov, and Welling}{Schlichtkrull et~al\mbox{.}}{2018}]%
        {schlichtkrull2018modeling}
\bibfield{author}{\bibinfo{person}{Michael Schlichtkrull},
  \bibinfo{person}{Thomas~N Kipf}, \bibinfo{person}{Peter Bloem},
  \bibinfo{person}{Rianne Van Den~Berg}, \bibinfo{person}{Ivan Titov}, {and}
  \bibinfo{person}{Max Welling}.} \bibinfo{year}{2018}\natexlab{}.
\newblock \showarticletitle{Modeling relational data with graph convolutional
  networks}. In \bibinfo{booktitle}{\emph{European Semantic Web Conference}}.
  Springer, \bibinfo{pages}{593--607}.
\newblock


\bibitem[\protect\citeauthoryear{Shi, Chen, Ma, Mao, Zhang, and Zhang}{Shi
  et~al\mbox{.}}{2020}]%
        {shi2020neural}
\bibfield{author}{\bibinfo{person}{Shaoyun Shi}, \bibinfo{person}{Hanxiong
  Chen}, \bibinfo{person}{Weizhi Ma}, \bibinfo{person}{Jiaxin Mao},
  \bibinfo{person}{Min Zhang}, {and} \bibinfo{person}{Yongfeng Zhang}.}
  \bibinfo{year}{2020}\natexlab{}.
\newblock \showarticletitle{Neural Logic Reasoning}. In
  \bibinfo{booktitle}{\emph{Proceedings of the 29th ACM International
  Conference on Information \& Knowledge Management}}.
  \bibinfo{pages}{1365--1374}.
\newblock


\bibitem[\protect\citeauthoryear{Sun, Deng, Nie, and Tang}{Sun
  et~al\mbox{.}}{2019}]%
        {sun2019rotate}
\bibfield{author}{\bibinfo{person}{Zhiqing Sun}, \bibinfo{person}{Zhi-Hong
  Deng}, \bibinfo{person}{Jian-Yun Nie}, {and} \bibinfo{person}{Jian Tang}.}
  \bibinfo{year}{2019}\natexlab{}.
\newblock \showarticletitle{Rotate: Knowledge graph embedding by relational
  rotation in complex space}.
\newblock \bibinfo{journal}{\emph{arXiv preprint arXiv:1902.10197}}
  (\bibinfo{year}{2019}).
\newblock


\bibitem[\protect\citeauthoryear{Toutanova, Chen, Pantel, Poon, Choudhury, and
  Gamon}{Toutanova et~al\mbox{.}}{2015}]%
        {toutanova2015representing}
\bibfield{author}{\bibinfo{person}{Kristina Toutanova}, \bibinfo{person}{Danqi
  Chen}, \bibinfo{person}{Patrick Pantel}, \bibinfo{person}{Hoifung Poon},
  \bibinfo{person}{Pallavi Choudhury}, {and} \bibinfo{person}{Michael Gamon}.}
  \bibinfo{year}{2015}\natexlab{}.
\newblock \showarticletitle{Representing text for joint embedding of text and
  knowledge bases}. In \bibinfo{booktitle}{\emph{Proceedings of the 2015
  conference on empirical methods in natural language processing}}.
  \bibinfo{pages}{1499--1509}.
\newblock


\bibitem[\protect\citeauthoryear{Trouillon, Welbl, Riedel, Gaussier, and
  Bouchard}{Trouillon et~al\mbox{.}}{2016}]%
        {trouillon2016complex}
\bibfield{author}{\bibinfo{person}{Th{\'e}o Trouillon},
  \bibinfo{person}{Johannes Welbl}, \bibinfo{person}{Sebastian Riedel},
  \bibinfo{person}{{\'E}ric Gaussier}, {and} \bibinfo{person}{Guillaume
  Bouchard}.} \bibinfo{year}{2016}\natexlab{}.
\newblock \showarticletitle{Complex embeddings for simple link prediction}.
  International Conference on Machine Learning (ICML).
\newblock


\bibitem[\protect\citeauthoryear{van~den Berg, Kipf, and Welling}{van~den Berg
  et~al\mbox{.}}{2018}]%
        {van2018graph}
\bibfield{author}{\bibinfo{person}{Rianne van~den Berg},
  \bibinfo{person}{Thomas~N Kipf}, {and} \bibinfo{person}{Max Welling}.}
  \bibinfo{year}{2018}\natexlab{}.
\newblock \showarticletitle{Graph Convolutional Matrix Completion}.
\newblock  (\bibinfo{year}{2018}).
\newblock


\bibitem[\protect\citeauthoryear{Veličković, Cucurull, Casanova, Romero,
  Liò, and Bengio}{Veličković et~al\mbox{.}}{2018}]%
        {veli2018graph}
\bibfield{author}{\bibinfo{person}{Petar Veličković},
  \bibinfo{person}{Guillem Cucurull}, \bibinfo{person}{Arantxa Casanova},
  \bibinfo{person}{Adriana Romero}, \bibinfo{person}{Pietro Liò}, {and}
  \bibinfo{person}{Yoshua Bengio}.} \bibinfo{year}{2018}\natexlab{}.
\newblock \showarticletitle{Graph Attention Networks}. In
  \bibinfo{booktitle}{\emph{International Conference on Learning
  Representations}}.
\newblock


\bibitem[\protect\citeauthoryear{Wang, Rong, Zhuo, and Zhu}{Wang
  et~al\mbox{.}}{2018}]%
        {wang2018embedding}
\bibfield{author}{\bibinfo{person}{Mengya Wang}, \bibinfo{person}{Erhu Rong},
  \bibinfo{person}{Hankui Zhuo}, {and} \bibinfo{person}{Huiling Zhu}.}
  \bibinfo{year}{2018}\natexlab{}.
\newblock \showarticletitle{Embedding knowledge graphs based on transitivity
  and asymmetry of rules}. In \bibinfo{booktitle}{\emph{Pacific-Asia Conference
  on Knowledge Discovery and Data Mining}}. Springer,
  \bibinfo{pages}{141--153}.
\newblock


\bibitem[\protect\citeauthoryear{Wang, Wang, and Guo}{Wang
  et~al\mbox{.}}{2015}]%
        {wang2015knowledge}
\bibfield{author}{\bibinfo{person}{Quan Wang}, \bibinfo{person}{Bin Wang},
  {and} \bibinfo{person}{Li Guo}.} \bibinfo{year}{2015}\natexlab{}.
\newblock \showarticletitle{Knowledge base completion using embeddings and
  rules}. In \bibinfo{booktitle}{\emph{Twenty-Fourth International Joint
  Conference on Artificial Intelligence}}.
\newblock


\bibitem[\protect\citeauthoryear{Wang, Wei, dos Santos, Wang, Nallapati,
  Arnold, Xiang, and Philip}{Wang et~al\mbox{.}}{2020}]%
        {wang2020h2kgat}
\bibfield{author}{\bibinfo{person}{Shen Wang}, \bibinfo{person}{Xiaokai Wei},
  \bibinfo{person}{Cicero dos Santos}, \bibinfo{person}{Zhiguo Wang},
  \bibinfo{person}{Ramesh Nallapati}, \bibinfo{person}{Andrew Arnold},
  \bibinfo{person}{Bing Xiang}, {and} \bibinfo{person}{S~Yu Philip}.}
  \bibinfo{year}{2020}\natexlab{}.
\newblock \showarticletitle{H2KGAT: Hierarchical Hyperbolic Knowledge Graph
  Attention Network}. In \bibinfo{booktitle}{\emph{Proceedings of the 2020
  Conference on Empirical Methods in Natural Language Processing (EMNLP)}}.
  \bibinfo{pages}{4952--4962}.
\newblock


\bibitem[\protect\citeauthoryear{Wang, He, Wang, Feng, and Chua}{Wang
  et~al\mbox{.}}{2019}]%
        {wang2019neural}
\bibfield{author}{\bibinfo{person}{Xiang Wang}, \bibinfo{person}{Xiangnan He},
  \bibinfo{person}{Meng Wang}, \bibinfo{person}{Fuli Feng}, {and}
  \bibinfo{person}{Tat-Seng Chua}.} \bibinfo{year}{2019}\natexlab{}.
\newblock \showarticletitle{Neural graph collaborative filtering}. In
  \bibinfo{booktitle}{\emph{Proceedings of the 42nd international ACM SIGIR
  conference on Research and development in Information Retrieval}}.
  \bibinfo{pages}{165--174}.
\newblock


\bibitem[\protect\citeauthoryear{Wang, Zhang, Feng, and Chen}{Wang
  et~al\mbox{.}}{2014}]%
        {wang2014knowledge}
\bibfield{author}{\bibinfo{person}{Zhen Wang}, \bibinfo{person}{Jianwen Zhang},
  \bibinfo{person}{Jianlin Feng}, {and} \bibinfo{person}{Zheng Chen}.}
  \bibinfo{year}{2014}\natexlab{}.
\newblock \showarticletitle{Knowledge graph embedding by translating on
  hyperplanes.}. In \bibinfo{booktitle}{\emph{AAAI}},
  Vol.~\bibinfo{volume}{14}. Citeseer, \bibinfo{pages}{1112--1119}.
\newblock


\bibitem[\protect\citeauthoryear{Yang, Yih, He, Gao, and Deng}{Yang
  et~al\mbox{.}}{2015}]%
        {yang2015embedding}
\bibfield{author}{\bibinfo{person}{Bishan Yang}, \bibinfo{person}{Scott Wen-tau
  Yih}, \bibinfo{person}{Xiaodong He}, \bibinfo{person}{Jianfeng Gao}, {and}
  \bibinfo{person}{Li Deng}.} \bibinfo{year}{2015}\natexlab{}.
\newblock \showarticletitle{Embedding Entities and Relations for Learning and
  Inference in Knowledge Bases}. In \bibinfo{booktitle}{\emph{Proceedings of
  the International Conference on Learning Representations (ICLR) 2015}}.
\newblock


\bibitem[\protect\citeauthoryear{Yang, Tian, Zhang, Yan, He, and Jin}{Yang
  et~al\mbox{.}}{2019}]%
        {yang2019transms}
\bibfield{author}{\bibinfo{person}{Shihui Yang}, \bibinfo{person}{Jidong Tian},
  \bibinfo{person}{Honglun Zhang}, \bibinfo{person}{Junchi Yan},
  \bibinfo{person}{Hao He}, {and} \bibinfo{person}{Yaohui Jin}.}
  \bibinfo{year}{2019}\natexlab{}.
\newblock \showarticletitle{TransMS: Knowledge Graph Embedding for Complex
  Relations by Multidirectional Semantics.}. In
  \bibinfo{booktitle}{\emph{IJCAI}}. \bibinfo{pages}{1935--1942}.
\newblock


\bibitem[\protect\citeauthoryear{Zhang, Paudel, Wang, Chen, Zhu, Zhang,
  Bernstein, and Chen}{Zhang et~al\mbox{.}}{2019}]%
        {zhang2019iteratively}
\bibfield{author}{\bibinfo{person}{Wen Zhang}, \bibinfo{person}{Bibek Paudel},
  \bibinfo{person}{Liang Wang}, \bibinfo{person}{Jiaoyan Chen},
  \bibinfo{person}{Hai Zhu}, \bibinfo{person}{Wei Zhang},
  \bibinfo{person}{Abraham Bernstein}, {and} \bibinfo{person}{Huajun Chen}.}
  \bibinfo{year}{2019}\natexlab{}.
\newblock \showarticletitle{Iteratively learning embeddings and rules for
  knowledge graph reasoning}. In \bibinfo{booktitle}{\emph{The World Wide Web
  Conference}}. \bibinfo{pages}{2366--2377}.
\newblock


\bibitem[\protect\citeauthoryear{Zhang, Chen, Yang, Ramamurthy, Li, Qi, and
  Song}{Zhang et~al\mbox{.}}{2020}]%
        {Zhang2020Efficient}
\bibfield{author}{\bibinfo{person}{Yuyu Zhang}, \bibinfo{person}{Xinshi Chen},
  \bibinfo{person}{Yuan Yang}, \bibinfo{person}{Arun Ramamurthy},
  \bibinfo{person}{Bo Li}, \bibinfo{person}{Yuan Qi}, {and} \bibinfo{person}{Le
  Song}.} \bibinfo{year}{2020}\natexlab{}.
\newblock \showarticletitle{Efficient Probabilistic Logic Reasoning with Graph
  Neural Networks}. In \bibinfo{booktitle}{\emph{ICLR}}.
\newblock


\end{thebibliography}

\newpage
\appendix
\section{Training Algorithm Pseudo-code}
\label{A:alg}
\begin{algorithm}[htb]
\caption{GCR Training Algorithm}
\SetKwInOut{Input}{Input}
\SetKwInOut{Output}{Output}
\SetAlgoLined
\Input{Graph $\mathcal{G(V, R, T)}$; triples $\mathcal{T}(h,r,t)\forall h,t\in \mathcal{V}, \forall r\in \mathcal{R}$; predicate function $P_r,\forall r\in \mathcal{R}$; epochs $K$; neighbor sample function $N$; negative sample function $S$; scoring function Sim; Graph Collaborative Reasoning network GCR; anchor vector $\mathbf{T}$; optimization algorithm $\mathrm{OPTIM}$; model parameters $\Theta$; logic regularizer weight $\lambda_{l}$; $\ell_2$ regularizer weight $\lambda_{\Theta}$; amplification coefficient $\alpha$}
% \Output{Latent logicalized vector representations $\mathbf{e}_v, \forall v\in \mathcal{V}$}
\BlankLine
Initialize node vectors $\mathbf{x}_v, \forall v\in \mathcal{V}$\;
Initialize predicate modules $P_r, \forall r\in \mathcal{R}$\;
% $\mathbf{e}_T^0 \leftarrow P_r(\mathbf{x}_h, \mathbf{x}_t), \quad \forall T\in \mathcal{T}, h,r,t\in T$\;
\For{epoch $k \leftarrow 1$ \KwTo $K$} {
    $\mathcal{L} \leftarrow 0$\;
    $\mathbf{e}_T^{k-1} \leftarrow P_r(\mathbf{x}_h, \mathbf{x}_t), \quad \forall T\in \mathcal{T}, h,r,t\in T$\;
    $\mathbf{e}_T^k \leftarrow \mathbf{e}_T^{k-1}/ ||\mathbf{e}_T^{k-1}||_2$\;
    \For{$T \in \mathcal{T}$} {
        $T' \leftarrow S(T)$\Comment{sample a fake triplet for $T$}\; 
        $\textbf{E} \leftarrow$ GCR($\mathbf{e}_T^{k}$, $\{\mathbf{e}_{T_N}^{k}, \forall T_N\in \mathcal{N}(T)\}$)\;
        $\textbf{E}' \leftarrow$ GCR($\mathbf{e}_{T'}^{k}$, $\{\mathbf{e}_{T_N}^{k}, \forall T_N\in \mathcal{N}(T)\}$)\;
        
        $s_T^+ \leftarrow$  Sim($\mathbf{E},\mathbf{T}$),~$s_{T'}^- \leftarrow$  Sim($\mathbf{E}',\mathbf{T}$)\;
        $\mathcal{L}_{gcr} \leftarrow -\ln \sigma (\alpha(s_T^+ - s_{T'}^-))$\;
        $\mathcal{L}_{logic} \leftarrow \sum_i r_i$ \Comment{logic constraints for logical laws}\;
        $\mathcal{L} \leftarrow \mathcal{L} + \mathcal{L}_{gcr} + \lambda_{l}\mathcal{L}_{logic} +\lambda_{\Theta}||\Theta||_2^2$\;
    }
    % $Loss \leftarrow Loss + \lambda_{\Theta}||\Theta||_2^2$\;
    OPTIM($\mathcal{L}$)\Comment{optimize all parameters for round $k$}\;
}
\end{algorithm}

\section{Experimental Settings}
\label{A:exp_setting}
\subsection{\textbf{Link Prediction}}
In the training stage, we first need to find the neighbors of the head and tail entity of the given triplet. Instead of using all the neighbor nodes to assemble the logical expression, we sample the neighbors uniformly, by following \cite{hamilton2017inductive}, in each iteration to predict the target triplet. 
%This can help to improve the training quality as well as the efficiency, since entities may have various numbers of neighbors and thus may result in very different lengths of the expression sequence. \textcolor{blue}{ADD CITATION}
% , the neural network may not be trained properly without any control to the length. 
In our implementation, we sample at most $n\in \{5, 10, 20\}$ neighbors for each entity in the given triplet. In other words, for each target triplet, the total number of neighbor triplets can be up to $2n$ ($n$ from the head entity and $n$ from the tail entity). To train the model, for each target triplet, we sample 1 negative triplet for pair-wise learning as mentioned in Eq.\eqref{eq:gln_loss}.

We set all vector embedding size 
% and the hidden output vector size 
to 64. The number of layers for predicate encoder networks and logical module networks is set to 3. The network parameters are initialized with normal distribution with mean 0 and standard deviation is 0.01. Dropout and $\ell_2$ regularization are adopted to avoid over-fitting. We set the dropout rate to 0.2 and the weight for $\ell_2$ regularizer $\lambda_\Theta$ is selected from $10^{-5}$ to $10^{-7}$. The logical regularizer weight $\lambda_l$ is selected in the range $10^{-1}$ to $10^{-7}$. We use Adam~\cite{kingma2014adam} as the optimization algorithm with learning rate initialized to 0.001 and learning rate decay is adopted during the training process. Early-stopping is used and the best model for reporting the results is selected based on the best performance on the validation set.

\subsection{\textbf{Recommendation}} For each user-item interaction in training set, we randomly sample the neighbors for both user and item nodes to construct the logical expression. We set the total number of neighbors for each user or item to 5, i.e. there will be at most 10 neighbor user-item interactions in the logical expression. We set the embedding size to 64 and the number of layers for network modules is 2. $\ell_2$ penalty weight $\lambda_{\Theta}$ is $10^{-5}$ for both datasets. The logical regularization weight $\lambda_{l}$ is $10^{-6}$. Learning rate is fixed at 0.001. Other settings are the same as the previous subsection.

For TransE, DistMult and ConvE, we set the embedding size to 100, while the embedding size and hidden size for BPR-MF and NCR are 64. $\ell_2$ weight for all baselines are $10^{-5}$. For ConvE, the number of channel is set to 32 and the kernel size is 3. For NCR, we use the open source implementation\footnote{\url{https://github.com/rutgerswiselab/NCR}}, more specifically, we apply the BPR-ranking loss to train the model and the neural logic modules have two layers with LeakyReLU as the activation function. Since NCR only considers nodes on user side, we only sample neighbor nodes on the user side. For NGCF, we also use the open source implementation in \cite{wang2019neural} to run the experiments. 

\section{Acknowledgement}
This work was supported in part by NSF IIS-1910154, IIS-2007907, IIS-2046457 and CCF-2124155. Any opinions and findings in this material are those of the authors and do not necessarily reflect those of the sponsors.

\end{document}